\documentclass[aps,pra,twocolumn,showpacs]{revtex4}
\usepackage{graphicx}

\usepackage{psfrag}

\usepackage{amsmath}

\usepackage{amssymb}

\renewcommand{\vec}[1]{\boldsymbol{#1}}

\renewcommand{\tensor}[1]{{\boldsymbol{\mathit {#1}}}}

\newcommand{\Nabla}{\vec{\nabla}}

\newcommand{\Lnabla}{\overset{\leftarrow}{\Nabla}}

\newcommand{\ket}[1]{\left| #1 \right\rangle}

\begin{document}
\title{Three-dimensional Casimir force between absorbing multilayer
dielectrics\footnote[1]
{Phys. Rev. A {\bfseries 68}, 033810 (2003),
Phys. Rev. A {\bfseries 69}, 019901(E) (2004)}
}
\author{Christian Raabe}
\author{Ludwig Kn\"{o}ll}
\author{Dirk-Gunnar Welsch}
\affiliation{Theoretisch-Physikalisches Institut,
Friedrich-Schiller-Universit\"at Jena, Max-Wien-Platz 1, D-07743 Jena,
Germany} 

\date{\today}

\begin{abstract}

Recently the
influence of dielectric and geometrical properties
on the Casimir force between dispersing and absorbing
multilayered plates in the zero-temperature limit has been
studied within a 1D quantization scheme for the electromagnetic
field in the presence of causal media 
[R. Esquivel-Sirvent, C. Villarreal, and G.H. Cocoletzi,
Phys. Rev. Lett. {\bf 64}, 052108 (2001)].
In the present paper a rigorous 3D analysis
is given, which shows that for complex heterostructures
the 1D theory only roughly reflects the dependence of
the Casimir force on the plate separation in general.  
Further, an extension of the very recently
derived formula for the Casimir force at zero
temperature [M.S. Toma\v{s}, Phys. Rev. A {\bfseries 66}, 052103
(2002)] to finite temperatures is given, and analytical
expressions for specific distance laws in the
zero-temperature limit are derived. In particular,
it is shown that 
the Casimir force between two single-slab plates behaves
asymptotically like $d^{-6}$ instead of $d^{-4}$
($d$, plate separation).  
\end{abstract}

\pacs{03.70.+k, 12.20.Ds, 42.60.Da}

\maketitle


\section{Introduction}
\label{sec1}

The Casimir effect has drawn attention to itself over
decades (for a review and a collection of references, see
\cite{Bordag,Lam}).
Various concepts and calculational techniques have been
developed, but only few of them have turned out to be capable of dealing
with realistic (at least Kramers-Kronig consistent) dielectric
bodies. The familiar `mode summation' method of deriving Casimir
forces as employed by Casimir himself \cite{Casimir}, for instance,
suffers from the obvious fact that there are no modes to expand
the (macroscopic) field operators when absorbing bodies are
involved rather than idealized boundaries. A generalized
mode decomposition (with the mode functions being orthogonal
with respect to an appropriate `weighted' inner product)
is appropriate only for (narrow) bandwidth-limited fields when dispersion
and absorption may be neglected \cite{WelschQuantumOptics}.
By contrast, all frequencies should be included in the calculation
of the Casimir force. Moreover, the
annoying infinities one has to cope with
when using modes always demand the application
of regularization techniques \cite{Bordag}.
To tackle realistic problems, one can
proceed in essentially three ways:
\begin{enumerate}
\item
The electromagnetic field and the material bodies are treated
macroscopically, and explicit field quantization is avoided
by invoking statistical thermodynamics to write down the field
correlation functions that are needed in the Maxwell stress tensor.
\item
The electromagnetic field and the material bodies are
quantized on a microscopic level. The bodies are described
by appropriate model systems that feature dissipation
of energy into a heat bath, and the coupled field-matter
equations are tried to be solved. In general, such a microscopic
approach to the problem is hardly feasible without rather sweeping
assumptions about the microscopic processes involved.
\item
The presence of dielectric bodies is described by means
of a spatially varying permittivity that is a complex function of
frequency, without addressing a specific microscopic model.
For given permittivity, the macroscopic, medium-assisted
electromagnetic field is quantized, by using a source-quantity
representation of the field in terms of the classical Green tensor
and an infinite set of appropriately chosen bosonic basic
fields.   
\end{enumerate}

The first method was introduced by Lifshitz \cite{Lifshitz}. He
obtained the force acting on two (semi-infinite)
absorbing dielectric walls by calculating the Maxwell stress tensor
in the region between the walls from field correlation functions,
which he could evaluate by employing a dissipation-fluctuation
relation for the `random electric field' that
acts as a Langevin noise source and balances the
effect of absorption in the Maxwell equations.
Although Lifshitz' calculation is not really quantum, he
was fully aware of the fact that the ``\ldots zero point
vibrations of the radiation field'' \cite{Lifshitz} cause the
effect at zero temperature. Lifshitz' results were
later re-derived by Schwinger \emph{et\,al.} using
source theory to circumvent quantization \cite{Schwinger}.
In a recent paper, Matloob \cite{Matloob} has essentially followed
Lifshitz, by postulating field correlation functions, without
explicit field quantization.      

An instructive treatment of the Casimir effect on the
basis of the second method was given by Kupiszewska \cite{Kup},
using a harmonic-oscillator model for the matter. The effect of
the heat bath accounting for dissipation was properly subsumed
within a Markovian damping term together with a
Langevin noise source corresponding to the random field in
Lifshitz' approach. Unfortunately, the calculations were carried out
only for one-dimensional (single-slab) systems. Recently,
this theory has been used to study the Casimir force between
multilayered plates \cite{MexicanGuys}. 

Here, we base the calculations on the third method
recently used by Toma\v{s} \cite{TomasCasimir}. It needs no external
input other than the (phenomenologically given) permittivities of
the bodies. Apart from dropping the bulk part of the Green tensor (giving rise
to unobservable bulk stress only), no regularization is needed.
Removal of the bulk stress, which is sometimes referred to as
``removing the Minkowski contribution'' in the literature,
is in fact necessary in all approaches, even in the framework of
Schwinger's clever source theory \cite{Schwinger}.

Let us briefly comment on two further techniques that
have been employed to calculate
Casimir forces. In the surface-mode approach 
originally used by van Kampen
\emph{et al.} \cite{vanKampen} to obtain Casimir forces in the
\emph{non-retarded limit}, the calculations are based on
a complete set of solutions
to Laplace's equation. Summing, by means of a complex
contour integral, the zero-point energy
assigned to each of these solutions
then yields the potential energy of the force. Later, the method was
extended to include
also retardation \cite{SurfaceModesRetarded}. However, in this
case \emph{all} normal modes and not only the surface
modes should  be considered. Though the method is
intrinsically based on mode expansion and hence on real
permittivities, correct results can also be found for absorbing material
if at some stage of the calculation the permittivities are allowed
to become complex. In the scattering approach \cite{ScatteringApproach},
which employs elastic (i.\,e. unitary) scattering
theory, the sum of the mode frequencies is
represented as an integral containing scattering phase shifts
or related quantities (see also \cite{Bordag}), and the
integration is performed in the complex plane. Again, if complex
permittivities are plugged into the resulting expressions
(derived for non-absorptive materials), correct
results can be obtained. Note that within
the framework of both methods the sum over the relevant mode
frequencies appears as the singularity contributions to
certain complex contour integrals, which are more generally valid 
than the initial expressions based on the
concept of normal modes.

In this paper, we first derive
a formula for the Casimir force between multilayer
dielectric plates at finite temperatures, which may be regarded
as being a generalization of the Lifshitz formula \cite{Lifshitz}
to multilayer systems and an extension to finite temperatures of
the zero-temperature result derived by Toma\v{s}  
\cite{TomasCasimir}. In particular for one-dimensional systems,
the formula reduces to an expression of the type derived by
Kupiszewska \cite{Kup}.
Transforming the formula to a form that is very suitable
for further analytical and numerical evaluation, we study the
dependence of the Casimir force on various system parameters
(such as the stacking order and the frequency response of the
permittivity) in detail. The numerical calculations are
performed for single-resonance dielectric matter of Drude-Lorentz
type. It is well known that the Casimir force between
semi-infinite dielectric walls
becomes proportional to $d^{-4}$ for large wall
separation $d$. We show that is also the `generic'
behavior for layered walls in general. However, yet different
types of long-distance laws are also possible.  
In particular, for single-slab walls of finite thickness, the
asymptotic $d^{-4}$ law changes to a $d^{-6}$ law.
We further show that the case of small
wall separation $d$ can be treated within Lifshitz'
approximations \cite{Lifshitz}.
   
The paper is organized as follows. In Section \ref{sec2} the
formalism is outlined and the basic formula for the Casimir force
is given. Section \ref{sec3} presents analytical results
for large and small wall separation, and Section \ref{sec4} is devoted
to the numerical results. A summary is
given in Section \ref{summary} followed by five appendices.
Appendix \ref{calculation} provides some basic relations
needed for the finite-temperature calculation.
Useful recurrence relations for the generalized reflection
coefficients are given in Appendix \ref{Fresnel}. The problem of switching to
imaginary frequencies in the basic integral expression for
the Casimir force is addressed in Appendix \ref{ContourFlip}.
In Appendix \ref{calculation1D} the formalism is applied
(for comparison) to one-dimensional systems, and in
Appendix \ref{Evaluation} a special integral is evaluated.


\section{Casimir force}
\label{sec2}

\subsection{Quantization scheme}
\label{sec2.1}

Let $\hat{\mathbf{E}}(\mathbf{r})$, $\hat{\mathbf{D}}(\mathbf{r})$,
$\hat{\mathbf{B}}(\mathbf{r})$, and $\hat{\mathbf{H}}(\mathbf{r})$
be the medium-assisted (ma\-cro\-scopic) electromagnetic field operators
such that
\begin{equation}
\label{2.0}
\hat{\mathbf{E}}(\mathbf{r})
=\int_0^\infty d\omega\,
\underline{\hat{\mathbf{E}}}(\mathbf{r},\omega)+ \mbox{H.c.},
\end{equation}
and $\hat{\mathbf{D}}(\mathbf{r})$, $\hat{\mathbf{B}}(\mathbf{r})$,
and $\hat{\mathbf{H}}(\mathbf{r})$ accordingly.
Within the framework of the quantization scheme given in
Ref.~\cite{Welsch},
the operators
$\underline{\hat{\mathbf{E}}}(\mathbf{r},\omega)$,
$\underline{\hat{\mathbf{D}}}(\mathbf{r},\omega)$,
$\underline{\hat{\mathbf{B}}}(\mathbf{r},\omega)$,
and $\underline{\hat{\mathbf{H}}}(\mathbf{r},\omega)$
obey Maxwell's equations 
\begin{align}
\label{2.2}
\Nabla\times\underline{\hat{\mathbf{E}}}(\mathbf{r},\omega)
-i\omega\underline{\hat{\mathbf{B}}}(\mathbf{r},\omega)=0,\\
\label{2.3}
\Nabla \underline{\hat{\mathbf{B}}}(\mathbf{r},\omega) =0,\\
\label{2.4}
\Nabla\times\underline{\hat{\mathbf{H}}}(\mathbf{r},\omega)
+i\omega\underline{\hat{\mathbf{D}}}(\mathbf{r},\omega) = 0,\\
\label{2.5}
\Nabla \underline{\hat{\mathbf{D}}}(\mathbf{r},\omega) =0,
\end{align}
where (for non-magnetic) linear media the constitutive
relations read as
\begin{gather}
\label{2.6}
\underline{\hat{\mathbf{D}}}(\mathbf{r},\omega)
=\varepsilon_0 \varepsilon(\mathbf{r},\omega)
\underline{\hat{\mathbf{E}}}(\mathbf{r},\omega)
+ \underline{\hat{\mathbf{P}}}_{\rm N}(\mathbf{r},\omega),\\
\label{2.7}
\underline{\hat{\mathbf{H}}}(\mathbf{r},\omega)
=\mu_0^{-1} \underline{\hat{\mathbf{B}}}(\mathbf{r},\omega).
\end{gather}
Here, the complex permittivity
\begin{equation}
\label{2.1}
\varepsilon(\mathbf {r},\omega)
= \varepsilon'(\mathbf {r},\omega) + i\varepsilon''(\mathbf {r},\omega)
\end{equation}
satisfies the Kramers-Kronig relations, and the noise polarization
\begin{equation}
\label{2.8}
\underline{\hat{\mathbf{P}}}_{\rm N}(\mathbf{r},\omega)
= i \sqrt{\hbar \varepsilon_0 \varepsilon''(\mathbf{r},\omega)/ \pi}
\,\hat{\mathbf{f}}(\mathbf{r},\omega)
\end{equation}
associated with material absorption is expressed in terms of 
bosonic fields $\hat{\mathbf{f}}(\mathbf{r},\omega)$, 
\begin{align}
\label{2.9}
[\hat{f}_i\mathbf{(r,\omega)},\hat{f}_j\mathbf{(r',\omega')}]&=0,
\\
\label{2.9a}
[\hat{f}_i\mathbf{(r,\omega)},\hat{f}^\dagger_j\mathbf{(r',\omega')}]
&=\delta_{ij}\delta\mathbf{(r-r')}\delta(\omega-\omega'),
\end{align}
which play the role of the dynamical variables of the
combined field-matter system consisting of the
electromagnetic field,
the medium polarization, and the heat bath
accounting for absorption.
From Eqs.~(\ref{2.2}) -- (\ref{2.8}) it then follows that
\begin{align}
\label{2.10}
\underline{\hat{\mathbf{B}}}(\mathbf{r},\omega)
&= (i\omega)^{-1}\Nabla \times
\underline{\hat{\mathbf{E}}}(\mathbf{r},\omega),\\
\label{2.11}
\underline{\hat{\mathbf{D}}}(\mathbf{r},\omega)
&= (\mu_0 \omega^2)^{-1}\Nabla \times \Nabla \times
\underline{\hat{\mathbf{E}}}(\mathbf{r},\omega),
\end{align}
where
\begin{equation}
\label{2.12}
\underline{\hat{\mathbf{E}}}(\mathbf{r},\omega)
= i \sqrt{\frac{\hbar}{\pi\varepsilon_0}}
\,\frac{\omega^2}{c^2} \!\int\! d^3r'\,
\sqrt{\varepsilon''(\mathbf{r}',\omega)}\,
\tensor{G}(\mathbf{r},\mathbf{r}',\omega)
 \hat{\mathbf{f}}(\mathbf{r}',\omega),
\end{equation} 
with $\tensor{G}(\mathbf{r},\mathbf{r}',\omega)$ being the
(classical) Green tensor, which is determined by the equation
\begin{equation}
\label{2.13}
\Nabla\times\Nabla\times\tensor{G}(\mathbf{r},\mathbf{r}',\omega)
-\frac{\omega^2}{c^2}\,\varepsilon(\mathbf{r},\omega)
\tensor{G}(\mathbf{r},\mathbf{r}',\omega)
= \tensor{\delta}(\mathbf{r},\mathbf{r}')
\end{equation}
[$\tensor{\delta}(\mathbf{r},\mathbf{r}')$, tensorial $\delta$-function]
together with `outgoing' boundary conditions. In particular,
the Green tensor satisfies the relations \cite{Welsch}
\begin{equation}
\label{2.13b}
G_{ij}({\mathbf r},{\mathbf r}',\omega)
= G_{ji}({\mathbf r}',{\mathbf r},\omega)
\end{equation}
and
\begin{equation}
\label{2.13c}
\frac{\omega^2}{c^2}\int d^3s\,
\varepsilon''(\mathbf{s},\omega)
\tensor{G}(\mathbf{r,s},\omega)
\tensor{G}^\ast(\mathbf{s,r'},\omega)
=\Im\!\left[\tensor{G}(\mathbf{r,r'},\omega)\right]
\end{equation}
($\Im$, imaginary part). The above given equations do not refer to a specific picture.
In particular, in the Heisenberg picture the operators
$\hat{\mathbf{f}}(\mathbf{r},\omega)$ simply carry an
exponential time dependence,
\begin{equation}
\label{2.13a}
\hat{\mathbf{f}}(\mathbf{r},\omega)
\mapsto \hat{\mathbf{f}}(\mathbf{r},\omega) e^{-i\omega t},
\end{equation} 
according to the Hamiltonian of the (overall) system
\begin{equation}
\label{2.20}
\hat{H}=\int_0^\infty d\omega \, \hbar\omega \int d^3r \,
\hat{\mathbf{f}}^{\dagger}(\mathbf{r},\omega)
\hat{\mathbf{f}}(\mathbf{r},\omega).
\end{equation}
%
\begin{figure}[ht]
\includegraphics[width=5cm]{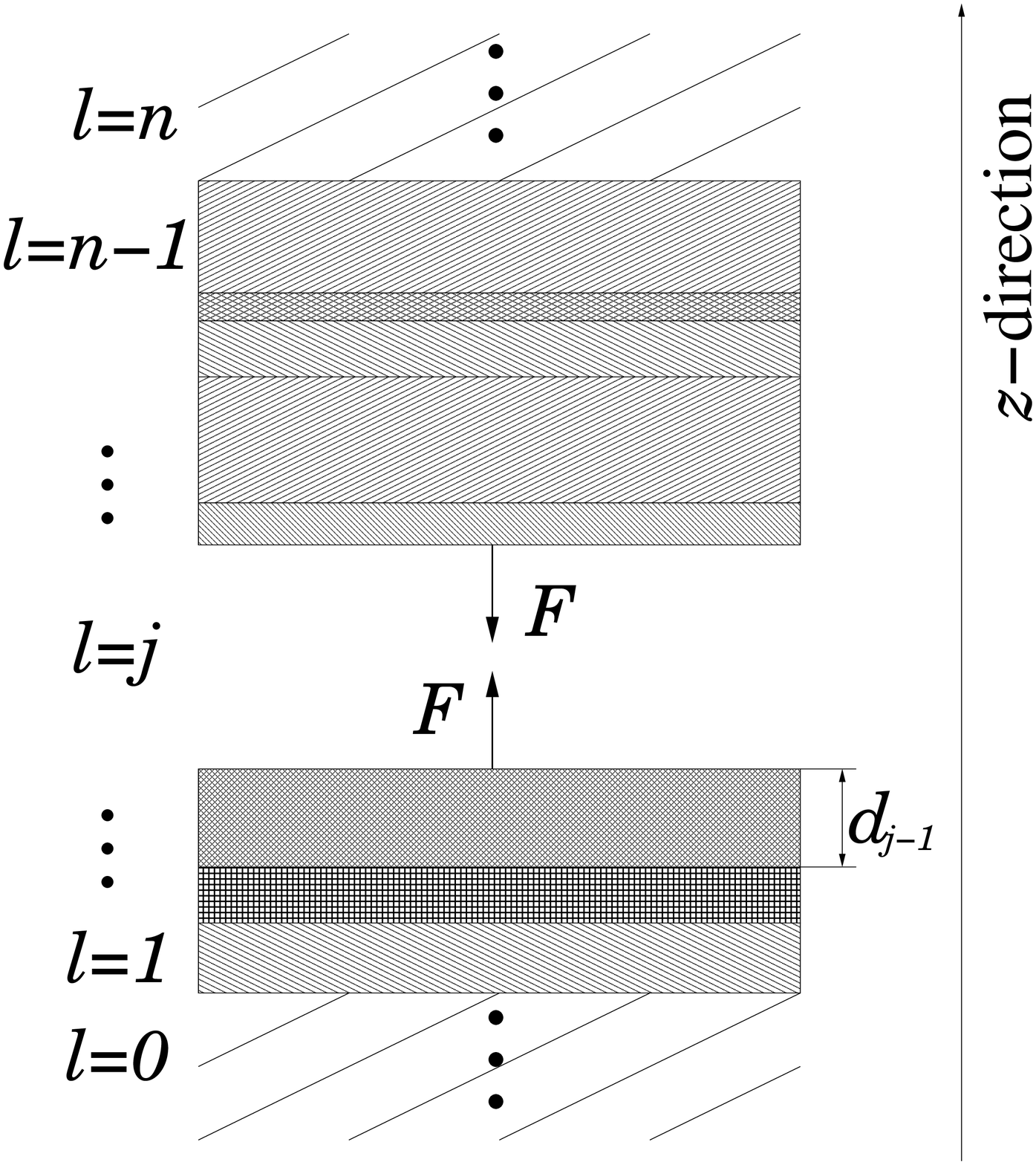}
\caption{\label{multilayer}
Scheme of the multilayer structure.}
\end{figure}%
For the multilayer structure sketched in
Fig.~\ref{multilayer} and considered throughout the paper,
\begin{equation}
\label{2.15}
\varepsilon(\mathbf{r},\omega)
=\varepsilon_l(\omega) \quad \mbox{for}\quad
z_l < z < z_{l+1}=z_l+d_{l}
\end{equation}
($l$ $\!=$ $\!0,\ldots,n$, $z_0$ $\!=$ $\!-\infty$,
$z_{n+1}$ $\!=$ $\!+\infty$), the Green tensor
$\tensor{G}_l(\mathbf{r},\mathbf{r}',\omega)$
in the $l$th layer can be decomposed as
\begin{equation}
\label{2.16}
\tensor{G}_l(\mathbf{r},\mathbf{r}',\omega)
=\tensor{G}_l^{\rm bulk}(\mathbf{r},\mathbf{r}',\omega)
+ \tensor{G}_l^{\rm scat}(\mathbf{r},\mathbf{r}',\omega),
\end{equation}
where $\tensor{G}_l^{\rm bulk}$ is the solution emerging from
a point-like source current placed in the $l$th layer 
without any boundaries, and $\tensor{G}_l^{\rm scat}$ solves
the homogeneous version of Eq.~(\ref{2.13}) so as to make
the full Green tensor obey the correct boundary conditions at the
surfaces of discontinuity. Since there is no Casimir effect
in homogeneous space, the bulk part can safely be dropped
(and actually \emph{must} be dropped) in
the stress tensor. The remaining scattering part contains the
geometrical data of the problem and is continuous at 
$\mathbf{r}$ $\!=$ $\!\mathbf{r}'$, whereas the bulk part is
rather singular there. The scattering part can be 
constructed as shown in Ref.~\cite{Tomas}. Here, only the case
where both spatial arguments are in the same layer is of interest. 


\subsection{Stress tensor}
\label{sec2.2}

In order to determine the Casimir force from the stress tensor,
we first calculate
\begin{multline}
\label{2.17}
\tensor{T}(\mathbf{r,r'},t)
= \tensor{T}_1(\mathbf{r,r'},t) + \tensor{T}_2(\mathbf{r,r'},t)
\\[1ex]
  - \textstyle{\frac{1}{2}}\,\tensor{I}\,
{\rm Tr}\,[\tensor{T}_1(\mathbf{r,r'},t)
+\tensor{T}_2(\mathbf{r,r'},t)],
\end{multline}
where
\begin{equation}
\label{2.18}
\tensor{T}_1\mathbf{(r,r'},t)
= \bigl\langle\hat{\mathbf{D}}(\mathbf{r},t) \otimes
\hat{\mathbf{E}}(\mathbf{r'},t)\bigr\rangle
\end{equation}
and
\begin{equation}
\label{2.19}
\tensor{T}_2\mathbf{(r,r'},t)
=\bigl\langle\hat{\mathbf{B}}(\mathbf{r},t) \otimes
\hat{\mathbf{H}}(\mathbf{r'},t)\bigr\rangle
\end{equation}
(\mbox{$\mathbf{r}$ $\!\neq$ $\!\mathbf{r}'$}).
Here, the electromagnetic-field operators
are thought of as being expressed in terms of the fundamental fields
$\hat{\mathbf{f}}(\mathbf{r},\omega)$ as outlined in
Section \ref{sec2.1}. To specify the quantum state,
we assume thermal equilibrium. 


\subsubsection{Basic equation}
\label{sec2.2.1}

For finite temperatures $T$, we may employ the canonical
density operator
\begin{equation}
\label{2.21}
\hat{\rho} = Z^{-1}
e^{-\hat{H}/(k_{\rm B}T)},
\end{equation}
where
\begin{equation}
\label{2.21a}
Z={\rm Tr}\, e^{- \hat{H}/(k_{\rm B}T)}.
\end{equation}
After some calculation we derive
(Appendix \ref{calculation}) 
\begin{eqnarray}
\label{2.22}
\lefteqn{
\tensor{T}_1\mathbf{(r,r')}
= \mbox{Tr}\,\bigl[\hat{\rho}\hat{\mathbf{D}}(\mathbf{r},t) \otimes
\hat{\mathbf{E}}(\mathbf{r'},t)\bigr]
}
\nonumber\\[1ex]&&\hspace{-1.5ex}
=\frac{\hbar}{\pi}\!\int_0^\infty \!\!d\omega
\,\coth\! \left(\!\frac{\hbar\omega}{2k_{\rm B} T}\!\right)
\!\frac{\omega^2}{c^2}\,
\Im\!\left[\varepsilon(\mathbf{r,\omega})
\tensor{G}(\mathbf{r,r',\omega})\right],
\qquad
\end{eqnarray}  
and
\begin{eqnarray}
\label{2.23}
\lefteqn{
\tensor{T}_2\mathbf{(r,r')}
= \mbox{Tr}\,\bigl[\hat{\rho}\hat{\mathbf{B}}(\mathbf{r},t) \otimes
\hat{\mathbf{H}}(\mathbf{r'},t)\bigr]
}
\nonumber\\[1ex]&&\hspace{-1.5ex}
=-\frac{\hbar}{\pi}\!\int_0^\infty\!\! d\omega
\,\coth\! \left(\!\frac{\hbar\omega}{2k_{\rm B} T}\!\right)
\Nabla\times
\Im\!\left[\tensor{G}(\mathbf{r,r',\omega})\right]
\!\times\!\Lnabla{'}.
\qquad
\end{eqnarray}    
Here and in the following the time argument $t$ is dropped (because
of stationarity). The multilayer Green tensor constructed in terms
of generalized Fresnel coefficients is given in Ref.~\cite{Tomas}.
The partial translational invariance of the problem (the layers
are assumed to have infinite lateral extension) naturally leads
to a decomposition of the Green tensor into an angular spectrum
of $s$- and $p$-polarized plane waves, whose wave vectors have
real components parallel to the multilayer surfaces.

Let the $j$th layer be free space. The Casimir force (per unit area)
between the two stacks separated by that
layer \footnote{The stress
   tensor is well-defined only in free-space
   regions. Therefore at the very end of the calculations
   the permittivity should be set equal to unity there.}
is then determined by the
$zz$-component of the stress tensor obtained from
$\tensor{T}\mathbf{(r,r'})$ in the coincidence
limit \mbox{$\mathbf{r'}$ $\!\to$ $\!\mathbf{r}$},
\begin{equation}
\label{2.24}
T_{zz,j}(\mathbf{r},\mathbf{r})
= \lim_{\mathbf{r}'\to\mathbf{r}}T_{zz,j}(\mathbf{r},\mathbf{r}')
\end{equation} 
(the index $j$ of a quantity indicates that the
quantity refers to the $j$th layer), where -- as already
mentioned -- the (divergent) bulk 
contribution to the Green tensor must be dropped.
Straightforward calculation yields
(\mbox{$\xi_\sigma$ $\!=$ $\!\delta_{\sigma p}$ $\!-$
$\!\delta_{\sigma s}$}, \mbox{$\sigma$ $\!=$ $\!s,p$},
\mbox{$M$ $\!=$ $\!\pm$}, \mbox{$N$ $\!=$ $\!\pm$}) 
\begin{eqnarray}
\label{2.25}
\lefteqn{
T_{zz,j}(\mathbf{r,r})
= -2\hbar \int_0^\infty d\omega \,
\coth\!\left(\frac{\hbar\omega}{2k_{\rm B} T}\right)
\ \times
}
\nonumber\\&&\hspace{-1ex}
\times\  
\Im\biggl[ \int_0^\infty \!\! dq \, q\beta_j^2(q,\omega) 
\!\!\sum_{M \ne N}\sum_\sigma 
\xi_\sigma g_{\sigma j}^{MN}(z,z,q,\omega)\biggr]\!. 
\qquad
\end{eqnarray}
Here, $\mathbf{q}$ is the transverse component of the wave vector
\mbox{$\mathbf{k}_j$ $\!=$ $\!(\mathbf{q},\beta_j)$}, whose
$z$-component (`propagation constant') $\beta_j$ is given by 
\begin{equation}
\label{2.26}
\beta_j=\beta_j(q,\omega)
=\sqrt{\omega^2\varepsilon_j(\omega)/c^2-q^2}\,,
\end{equation}
and
\begin{equation}
\label{2.27}
g_{\sigma j}^{MN}(z,z',q,\omega)
\sim\exp {(M\,i\beta_{j}z)}\exp{(N\,i\beta_{j} z')}
\end{equation}
is related to the scattering part of the Green tensor
as \mbox{[$\mathbf{r}$ $\!=$ $\!(\vec{\rho},z)$]}
\begin{eqnarray}
\label{2.28}
&&\hspace{-5ex}
\tensor{G}_j^{\rm scat}(\mathbf{r},\mathbf{r'} ,\omega)
=\int d^2q \,e^{i\mathbf{q}(\vec{\rho}-\vec{\rho'})}
\tensor{G}_j^{\rm scat}(z,z',\mathbf{q},\omega),
\\
&&\hspace{-5ex}
\tensor{G}_j^{\rm scat}(z,z',\mathbf{q},\omega)
\nonumber\\
\label{2.29}
&&\hspace{-3ex}
= \sum_{\sigma} \sum_{M,N}
g_{\sigma j}^{MN}(z,z',q,\omega)
\mathbf{{e}}_{\sigma j}^M(\mathbf{q})\otimes
\mathbf{{e}}_{\sigma j}^N(\mathbf{-q}),
\quad
\end{eqnarray}
with polarization unit vectors 
\begin{equation}
\label{2.29a}
\mathbf{{e}}_{s j}^{\pm}(\mathbf{q}) = \frac{\mathbf{q}}{q}\times
\mathbf{e}_{z},\quad
\mathbf{e}_{pj}^{\pm}(\mathbf{q})=
\frac{1}{k_j} 
\left(q\mathbf{e}_{z}\mp\beta_{j}\frac{\mathbf{q}}{q}\right) 
\end{equation}   
[$\mathbf{e}_{z}$, unit vector in $z$-direction]. Making use of the explicit form of the
Green tensor as given in \cite{Tomas}, we find after some algebra
\begin{multline}
\label{2.30}
\beta_j^2\sum_{M
\ne N}\sum_\sigma \xi_\sigma g_{\sigma j}^{MN}
\\
= \frac{i \beta_j}{4\pi^2}\,e^{2i\beta_j d_{j}}
\sum_\sigma D_{\sigma j}^{-1}r_{j-}^\sigma r_{j+}^\sigma \cos[\beta_j (z-z')],
\end{multline} 
where
\begin{equation}
\label{2.31}
D_{\sigma j}=D_{\sigma j}(q,\omega)
=1-r_{j+}^\sigma r_{j-}^\sigma e^{2i\beta_j d_{j}}
\end{equation} 
can be thought of as accounting for multiple
reflections, with the
generalized Fresnel coefficients \mbox{$r_{j\pm}^\sigma$ $\!=$
$\!r_{j\pm}^\sigma(q,\omega)$} being the reflection
coefficients for $\sigma-$polarized waves at the top ($+$)
and bottom ($-$) of the $j$th layer.
They can be calculated recursively (for useful recurrence relations, see
Appendix \ref{Fresnel}) and in this way expressed in
terms of the thicknesses and permittivities of the layers
that are actually under consideration. 
Inserting Eq.~(\ref{2.30}) into Eq.~(\ref{2.25})
eventually yields
\begin{multline}
\label{2.32}
T_{zz,j}=
-\frac{\hbar}{2\pi^2}\int_0^\infty
d\omega\, \coth \!\left(\frac{\hbar\omega}{2k_{\rm B}T}\right)
\ \times
\\
\ \times
\Re\biggl(\int_0^\infty dq \,q \beta_j e^{2i\beta_j d_{j}}
\sum_\sigma D_{\sigma j}^{-1} r_{j-}^\sigma r_{j+}^{\sigma}\biggr)
\end{multline}
($\Re$, real part). Since $T_{zz,j}$ does not depend on the space point in the
$j$th layer, the argument $\mathbf{r}$ has been dropped, and
Eq.~(\ref{2.32}) gives the Casimir force (per unit area)
that acts on arbitrary multilayered walls.

It should be pointed out that in Eq.~(\ref{2.32})
nothing is said about the details of stratification of the walls.
In fact, any stratified system (whose material properties change only along
one direction -- here the $z$-direction)
admits of separation into an angular
spectrum of $s$- and $p$-polarized fields which do never mix
\cite{BornWolf}. This implies that Eq.~(\ref{2.32}),
where the Casimir force is expressed in terms of the reflection
coefficients of the walls, is very general and also applies 
to walls whose permittivity is a (on a macroscopic
scale) continuously varying function of $z$. For such walls, however,
the reflection coefficients
cannot be calculated from simple recurrence relations. In any case,
they might be determined experimentally. 

When in the zero-temperature limit the overall system is in
the ground state $\ket{0}$, so that
\mbox{$\hat{\mathbf{f}}(\mathbf{r},\omega)\ket{0}$ $\!=0$
$\forall \: \mathbf{r},\omega$}, then the thermal weighting factor
$\coth [\hbar\omega/(2k_{\rm B}T)]$ does not appear
in Eqs.~(\ref{2.22}) and (\ref{2.23}) and in the equations that
follow from them, and thus Eq.~(\ref{2.32}) changes to
Toma\v{s}' formula \cite{TomasCasimir}
\begin{eqnarray}
\label{2.35}
\lefteqn{
T_{zz,j}
=\,-\frac{\hbar}{2\pi^2} \!\int_0^\infty \!\!d\omega\,
\ \times
}
\nonumber\\&&
\times\ 
\Re\biggl(\int_0^\infty dq \,q
\beta_j e^{2i\beta_j d_{j}}\sum_\sigma D_{\sigma j}^{-1}
r_{j-}^\sigma r_{j+}^{\sigma}\biggr).
\end{eqnarray}
%

\subsubsection{Imaginary frequencies}
\label{sec2.2.2}

Exploiting the analytical properties of the $\omega$-integrand
in Eq.~(\ref{2.32}) in the upper complex frequency half-plane, we can
equivalently rewrite (\ref{2.32}) as
\begin{multline}
\label{2.36}
T_{zz,j}=
\frac{\hbar}{2\pi^2}\lim_{\eta\to 0+}
\Im\Biggl\{ \int_0^\infty
d\xi\, \biggl[\coth \!\left(\frac{\hbar\omega}{2k_{\rm B}T}\right)
\ \times
\\
\ \times
\int_0^\infty dq \,q \beta_j e^{2i\beta_j d_{j}}
\sum_\sigma D_{\sigma j}^{-1} r_{j-}^\sigma r_{j+}^{\sigma}
\biggr]_{\omega=\eta+i\xi}
\Biggr\},
\end{multline}
i.e. the frequency integration is performed on a straight line
parallel (and infinitesimally close) to the imaginary frequency axis
(for details, see Appendix \ref{ContourFlip}). Note
that (the small) $\eta$ in Eq.~(\ref{2.36})
only indicates that the (first-order) poles of the hyperbolic
cotangent
at the imaginary frequencies
\begin{equation}
\label{2.37}
\omega_m=i\xi_m=2i m\pi k_{\rm B} T/\hbar
\end{equation}
($m$, integer)
have to be kept to the left of the integration contour.

In order to further evaluate $T_{zz,j}$ as given by
Eq.~(\ref{2.36}), we first note that in the limit
\mbox{$\eta$ $\!\to$ $\!0+$}
the hyperbolic cotangent becomes purely imaginary.
Since the permittivity is purely real and
positive on the (positive!) imaginary frequency axis (see, e.g.,
\cite{LanLif}), the propagation constant
becomes purely imaginary,
\begin{equation}
\label{2.38}
\beta_j(q,\omega=i\xi)=i\kappa_j=i\sqrt{\xi^2\varepsilon_j(i\xi)/c^2+q^2},
\end{equation}
and thus the generalized reflection coefficients become real.
Hence, the intervals between
the poles (\ref{2.37}) do not contribute 
to the imaginary part in Eq.~(\ref{2.36}), since the term
within the square bracket becomes purely real. Clearly, the same
conclusions can also be drawn  
directly from Eqs.~(\ref{2.22}) and
(\ref{2.23}),  because the imaginary parts of both
the permittivity and the Green tensor vanish at imaginary
frequencies. In this way, the $\xi$-integral in Eq.~(\ref{2.36}) can be
given by a residue series according to
\mbox{($\eta$ $\!\to$ $\!0+$)}
\begin{multline}
\label{2.39}
\Im\biggl[\int_0^\infty d\xi\,
f(i\xi)
\biggr]\\
=2\pi\Im
\biggl[
{\textstyle\frac{1}{2}}
\sum_{m=0}^\infty \left(1-{\textstyle\frac{1}{2}}\delta_{m0}
\right)
\mbox{Res}\,f(\omega_m)
\biggr], 
\end{multline}
with the poles from (\ref{2.37}). Taking into account that
\begin{multline}
\label{2.40}
\lim_{\omega\to\omega_m}(\omega-\omega_m)
\coth\!\left(\frac{\hbar\omega}{2k_{\rm B}T}\right)
\\
=\left[\frac{\partial}{\partial\omega_m}
\tanh\!\left(\frac{\hbar\omega_m}{2k_{\rm B}T}\right)\right]^{-1}
=2k_{\rm B} T/\hbar, 
\end{multline}
[the rest of the integrand is holomorphic; see
Appendix \ref{ContourFlip}], we eventually derive
\begin{multline}
\label{2.41}
T_{zz,j}=\frac{ k_{\rm B} T}{\pi}
\sum_{m=0}^\infty
\left(1-{\textstyle\frac{1}{2}}\delta_{m0}\right)
\ \times\\ \times\ 
\biggl[\int_0^\infty dq \,q \kappa_j e^{-2\kappa_j d_{j}}
\sum_\sigma D_{\sigma j}^{-1} r_{j-}^\sigma
r_{j+}^{\sigma}\biggr]_{\omega=i\xi_m}\!,
\end{multline}
which may be regarded as a generalization of the popular
Lifshitz formula \cite{Lifshitz}. Note that
the zero-frequency term in the Lifshitz formula has been
a subject of controversial debate,
because the reflection coefficients can be discontinuous
at the point $(q$ $\!=$ $\!0$, $\!\omega$ $\!=$ $\!0)$, so that
it matters from which direction this point is approached
\footnote{Moreover, Lifshitz' original variables were generally
criticized for being singular, see e.g. \cite{Schwinger}.}.
From the above given derivation of Eq.~(\ref{2.41}) it
follows that first the 
$q$-integral for a small but non-zero
value of $\xi_0$ should be calculated and then
one can let \mbox{$\xi_0$ $\!\to$ $\!0+$}. It should be mentioned 
that the feasibility of flipping the contour of the frequency
integration results from the
properties of the Green tensor in the
\emph{position space}, not in the $(\mathbf{q},z)$-space.
      
To obtain $T_{zz,j}$ in the zero-temperature limit, we may
simply set (thanks to $\eta$) \mbox{$T$ $\!=$ $\!0$} in
Eq.~(\ref{2.36}). The thermal weighting factor thus reduces
to unity and we have
\begin{multline}
\label{2.42}
T_{zz,j}
= \frac{\hbar}{2\pi^2}
\Im \Biggl\{\int_0^\infty\!\!
d\xi  \!\int_0^\infty \!\!dq \,q
\ \times\\ \times\ 
\biggl[\beta_j e^{2i\beta_j d_{j}}
\!\sum_\sigma D_{\sigma j}^{-1} r_{j-}^\sigma
r_{j+}^{\sigma}\biggr]_{\omega=i\xi}
\Biggr\},
\end{multline}
which we may rewrite, on using Eqs.~(\ref{2.31})
and (\ref{2.38}), as 
\begin{multline}
\label{2.43}
T_{zz,j}
= \frac{\hbar c}{2\pi^2} \!\int_0^\infty\!\!
\frac{d\xi}{c}
\!\int_0^\infty \!\! dq \,q\kappa_j e^{-2\kappa_j d_{j}}
\ \times\\ \times\ 
\!\biggl[\sum_\sigma \!\frac{r_{j-}^\sigma
r_{j+}^{\sigma}}{1-r_{j-}^\sigma r_{j+}^\sigma
e^{-2\kappa_j d_{j}}}\!\biggr]_{\omega=i\xi}\!.
\end{multline}
Of course, Eq.~(\ref{2.43}) can be obtained by replacing
$\sum_{m=0}^\infty \cdots$ with $\hbar\,(2\pi k_{\rm B}
T)^{-1}\int_0^\infty d\xi \cdots$ in Eq.~(\ref{2.41}), since the
distance \mbox{$\Delta\xi_m$ $\!=$ $\!2\pi k_{\rm B} T/\hbar$} between
neighboring poles [see Eq.~(\ref{2.37})]
becomes small in the zero-temperature limit, or, 
alternatively, by flipping the frequency integration contour
into the imaginary axis directly in Eq.~(\ref{2.35}) \cite{TomasCasimir}. 
\subsubsection{1D systems}
\label{sec2.2.3}

In order to compare the above given macroscopic approach to the Casimir force
with the more microscopic approach developed in \cite{Kup}
for one-dimensional systems, we have to perform
our calculations for one field component only, e.g.,
\begin{equation}
\label{2.45}
\underline{\hat{\mathbf{E}}}(\mathbf{r},\omega)
= {\cal A}^{-1/2} \underline{\hat{E}}(z,\omega) \mathbf{e}_x
\end{equation}
(${\cal A}$, normalization area;
$\mathbf{e}_x$, unit vector in $x$-direction).
According to the quantization scheme outlined in
Sections \ref{sec2.1},
the (effectively) scalar electric field strength
$\underline{\hat{E}}(z,\omega)$
can be expressed in terms
of scalar bosonic basic fields and a scalar Green function. 
Following the line in Sections
\ref{sec2.1} and \ref{sec2.2}
and restricting, for simplicity, our attention to the
zero-temperature limit, we derive (Appendix \ref{calculation1D}) 
\begin{equation}
\label{2.46}
T_{zz,j}=-\frac{\hbar}{\pi\mathcal{A}}\int_0^\infty d\omega
\,\Re\biggl[\frac{\beta_j r_{j+}r_{j-}e^{2i\beta_jd_{j}}}{D_j}
\biggr],
\end{equation}
with $\mathcal{A}$ $\!(\to$ $\!\infty)$
being the normalization area
perpendicular to the $z$-direction. Note that instead
of Eq.~(\ref{2.26}) now 
\begin{equation}
\label{2.47}
\beta_j=\beta_j(\omega)
= \frac{\omega}{c}\,\sqrt{\varepsilon_j(\omega)}\,
\end{equation}
is valid, and the polarization index
$\sigma$ can be omitted,
since Eq.~(\ref{2.45}) implies normal
incidence and fixed polarization.
Equation (\ref{2.46}) can formally be obtained from
Eq.~(\ref{2.35}) by making the replacement   
\begin{equation}
\label{2.48}
\frac{1}{4\pi^2}\int d^2q
\quad\mapsto\quad
\frac{1}{\mathcal{A}}\sum_{\mathbf{q}}
\end{equation}
and keeping only the normally incident waves 
\mbox{($\mathbf{q}$ $\!=$ $\!0$)} of fixed ($s$ or $p$)
polarization $\sigma$.

In the case of two identical plates 
(separated by vacuum,
i.e. \mbox{$\varepsilon_j$ $\!=$ $\!1$}), the reflection
coefficients $r_{j+}$ and $r_{j-}$ can be identified with
the single-plate reflection coefficient $r_j$, i.e.
\mbox{$r_{j\pm}$ $\!=$ $\!r_j$}. From Eq.~(\ref{2.46})
it then follows that the Casimir force
(per unit area) resulting from the
two possible polarizations, \mbox{$F_{1D}$ $\!=$
$\!2 T_{zz,j}$}, can be given by
\begin{equation}
\label{2.49}
F_{1D} 
=-\frac{2\hbar}{\pi c{\cal A}}\int_0^\infty d\omega\,\omega
\Re\biggl[ \frac{r_j^2 e^{2i\omega d_{j}/c}}{1-r_j^2
\exp{(2i\omega d_{j}/c)}}
\biggr],
\end{equation}
which corresponds to the result obtained in \cite{Kup},
if the exact permittivity of the plate material is identified
with the model permittivity in \cite{Kup}. 
Note that $F_{1D}{\cal A}$ is the total force acting on the normalization area
\mbox{$\mathcal{A}$ $\!(\to$ $\!\infty)$} and not the force
per unit area (as erroneously stated in \cite{MexicanGuys}).  


\section{Asymptotic distance laws}
\label{sec3}
 
Let us consider the zero-temperature limit in more detail.
In order to evaluate
Eq.~(\ref{2.43}) for the case when
\mbox{$\varepsilon_j$ $\!=$ $\!1$} is valid, it is appropriate
to change to polar coordinates,
\begin{equation}
\label{2.50}
\xi/c=\kappa_{j} \cos \phi,\quad q=\kappa_{j} \sin \phi,
\end{equation}
and rewrite Eq.~(\ref{2.43}) as
\begin{multline}
\label{2.51}
F \equiv T_{zz,j} =\frac{\hbar c}{2\pi^2} \int_0^\infty
d\kappa_{j} \,\kappa_{j}^3 e^{-2\kappa_{j} d_{j}}
\int_0^{\pi/2} d\phi\, \sin\phi
\ \times\\ \times\ 
\sum_\sigma \frac{r_{j-}^\sigma
r_{j+}^{\sigma}}{1-r_{j-}^\sigma r_{j+}^\sigma \exp(-2\kappa_{j} d_{j})}\,.
\end{multline}
Here and in the following we do not explicitly indicate
that \mbox{$r^\sigma_{j\pm}$ $\!=$
$\!r^\sigma_{j\pm}(\omega$ $\!=$ $\!ic\kappa_{j}\cos\phi,
q$ $\!=$ $\!\kappa_{j}\sin\phi)$}.
Now, making the $\kappa_{j}$-integral dimensionless by letting
\begin{equation}
\label{2.52}
u=e^{-2\kappa_{j} d_{j}},
\quad
du=-2d_{j}\, e^{-2\kappa_{j} d_{j}} d\kappa_{j} 
\end{equation}
at once extracts the (generic) asymptotic dependence
on $d_{j}$ and also yields a finite domain
of integration (recommended for numerical computations)
\begin{multline}
\label{2.53}
F= - F_0(d_{j})
\ \times\\ \times\ 
\frac{15}{2 \pi^4} \!\int_{0+}^1\!\!
du \,\ln^3\!u \int_0^{\pi/2}\!\! d\phi \, \sin\phi
\sum_\sigma \frac{r_{j-}^\sigma
r_{j+}^{\sigma}}{1-r_{j-}^\sigma r_{j+}^\sigma u}\,,
\end{multline}
where
\begin{equation}
\label{2.54}
F_0(d_{j})= \frac{\hbar c \pi^2}{240}\frac{1}{d_{j}^4}
\end{equation}
is the
well-known formula for
the force (per unit area)
between two perfectly reflecting walls,
which was first derived by Casimir \cite{Casimir}.

\subsection{Standard long-distance law}
\label{sec3.1}

Appendix \ref{Fresnel} shows that the reflection coefficients
$r_{j\pm}^\sigma$ as functions of $q$ and $\omega$, 
\mbox{$r_{j\pm}^\sigma$ $\!=$ $\!r_{j\pm}^\sigma(q,\omega)$},
do not depend on the distance $d_{j}$ between the two reflecting
walls. Any dependence on $d_{j}$ of the integral expression in
Eq.~(\ref{2.53}) is therefore a matter of how the reflection
coefficients scale with the variable $u$. If $d_{j}$ is large,
then only the values
of $r_{j\pm}^\sigma$ for sufficiently large wavelengths
(i.e. small values of both $q$ and $\xi$)
can effectively contribute to the integral expression
in Eq.~(\ref{2.53}), whereas the other ones are
exponentially suppressed.

Introducing the `static' values of the $\phi$-integrals
of $(r_{j+}^\sigma r_{j-}^\sigma)^m$,
\begin{equation}
\label{2.54a}
\overline{\left(r_{j+}^\sigma r_{j-}^\sigma\right)^m}
=\lim_{\kappa_{j}\to 0+}\int_0^{\pi/2} d\phi\, 
\sin \phi\, \left(r_{j+}^\sigma r_{j-}^{\sigma}\right)^m,
\end{equation}
we may evaluate Eq.~(\ref{2.51}) [or Eq.~(\ref{2.53})] in the
large-distance limit to obtain
\begin{equation}
\label{2.55}
F=\frac{F_0(d_{j})}{2 \zeta(4)}\sum_\sigma
\overline{{\rm Li}_4\!\left({r_{j+}^\sigma r_{j-}^\sigma}\right)}
\qquad (d_{j}\to\infty),
\end{equation}
where we have expanded
\mbox{$1/[1$ $\!-$ $\!r_{j+}^\sigma r_{j-}^\sigma
e^{-2\kappa_{j} d_{j}}]$} in powers of $r_{j+}^\sigma r_{j-}^\sigma e^{-2\kappa_{j} d_{j}}$
and replaced the resulting $\phi$-integrals of
powers of $r_{j+}^\sigma r_{j-}^\sigma$ by their `static' values
according to Eq.~(\ref{2.54a}). In Eq.~(\ref{2.55}), 
${\rm Li}_s(z)$ $\!=$ $\!\sum_{m=1}^\infty z^m/m^s$
is the polylogarithm function (the series converges for
\mbox{$|z|$ $\!\le$ $\!1$}, \mbox{$s$ $\!>$ $\!1$}), and
$\zeta(s)$ $\!=$ $\!\sum_{m=1}^\infty m^{-s}$ is the
Riemann zeta function \mbox{[$\zeta(4)$ $\!=$ $\!\pi^4/90$]}.
Clearly, Eq.~(\ref{2.55}) gives the correct asymptotics
only if the `static' polylogarithm does not vanish for
the two polarizations (see Section \ref{sec3.2}).
Using the relation
\begin{equation}
\label{2.56}
{\rm Li}_s(|x| \leq 1) \leq \zeta(s) = {\rm Li}_s( x \to 1-)
\qquad (s>1),
\end{equation}
from Eq.~(\ref{2.55}) we see
that the asymptotic value of the force is bounded by
$F_0(d_{j})$. Note that assuming constant reflection coefficients would formally
produce the well-known $d_{j}^{-4}$ distance law for arbitrary $d_{j}$.
However, this unphysical assumption clearly contradicts the
validity of Eq.~(\ref{2.51}).

Let us briefly discuss the validity of Eq.~(\ref{2.55}).
In order to replace $(r_{j+}^\sigma r_{j-}^\sigma)^m$ by
$\overline{(r_{j+}^\sigma r_{j-}^\sigma)^m}$ according to Eq.~(\ref{2.54a}),
the reflection coefficients
$r_{j\pm}^\sigma$ as functions of $\kappa_{j}$ must be slowly
varying on a $\kappa_{j}$-scale of the order of magnitude
of $d_{j}^{-1}$. From the structure of the reflection coefficients
(Appendix \ref{Fresnel}) it is seen that they depend
on $\kappa_{j}$ via the (dependence on frequency of the)
permittivities $\varepsilon_l$ of the layers and
the exponentials $\exp(-2\kappa_l d_{l})$. Hence, two
conditions must be satisfied. If $\xi_l$ is the
characteristic frequency scale of variation
of $\varepsilon_l$ on the imaginary frequency axis,
then $\kappa_{j}$ $\!\approx$ $\!d_{j}^{-1}$
must be small compared with $\xi_l/c$. Thus, one
condition can be given by 
\begin{equation}
\label{2.57}
d_{j} \gg \frac{c}{\xi_l}  
\end{equation}
\mbox{($l$ $\!\neq$ $\!j$)}, i.e. the characteristic
wavelength $d_{j}$ of the `cavity' formed
by the two multilayered walls must be much
larger than the characteristic wavelengths of all the wall
permittivities \footnote{Typically, $\xi_l$ is related
to the lowest resonance frequency of a dielectric layer or
the plasma frequency of a metallic layer.}.
The other condition comes from the requirement
that \mbox{$\kappa_l d_{l}$ $\!\ll$ $\!1$} on the relevant
$\kappa_{j}$-scale mentioned above. Recalling Eqs.~(\ref{2.38}) and (\ref{2.50}) and
the condition (\ref{2.57}), we thus arrive at the
condition that
\begin{equation}
\label{2.60}
d_{j} \gg \sqrt{\varepsilon_l(i\xi_l)}\, d_{l} 
\end{equation}
\mbox{($l$ $\!\neq$ $\!0,j,n$)}. Note that in the case of
semi-infinite walls considered by Lifshitz \cite{Lifshitz} only the condition
(\ref{2.57}) is needed.


\subsection{Non-standard long-distance laws}
\label{sec3.2}

Let us consider, e.g., the case when
for small values of $\kappa_{j}$ the relation
\begin{equation}
\label{2.61}
r_{j\pm}^\sigma
\simeq
\kappa_{j} R_{j\pm}^\sigma(\phi)
\end{equation}
is valid, with the $R_{j\pm}^\sigma$ being bounded functions
of $\phi$, so that we may write
\begin{equation}
\label{2.62}
\frac{r_{j-}^\sigma
r_{j+}^{\sigma}}{1-r_{j-}^\sigma r_{j+}^\sigma \exp(-2\kappa_{j} d_{j})}
= \kappa_{j}^2 R_{j-}^\sigma R_{j+}^{\sigma} + {\cal O}(\kappa_{j}^4)
\end{equation}
We substitute this expression into Eq.~(\ref{2.51}) and
find that the leading term of the force in the large-distance
limit now reads as

\begin{eqnarray}
\label{2.63}
F&=&\frac{\hbar c}{2\pi^2}\sum_\sigma \int_0^\infty
d\kappa_{j}\, \kappa_{j}^{5}  e^{-2\kappa_{j} d_{j}}
\,\overline{R_{j-}^\sigma R_{j+}^\sigma}\nonumber\\
&=&\frac{15 \hbar c}{16 \pi^2
d_{j}^6}\sum_\sigma \overline{R_{j-}^\sigma R_{j+}^\sigma}
\qquad (d_{j}\to\infty),
\end{eqnarray}
where
\begin{equation}
\label{2.64}
\overline{R_{j-}^\sigma R_{j+}^\sigma}
=\lim_{\kappa_{j} \to 0+}\kappa_{j}^{-2}\int_0^{\pi/2}d\phi\, \sin \phi\,
r_{j-}^\sigma r_{j+}^\sigma. 
\end{equation}
Thus, the Casimir force asymptotically behaves like
\mbox{$\sim$ $\!d_{j}^{-6}$}.
Obviously, other non-standard large-distance laws can also
be observed. In particular, when the relation (\ref{2.61}) is valid
for either $r_{j-}^\sigma$ or $r_{j+}^\sigma$, then the
Casimir force asymptotically behaves like \mbox{$\sim$ $\!d_{j}^{-5}$}.  

To be more specific, let us consider the reflection
coefficients $r_{j\pm}^s$
in more detail. By assuming finite values of
$\varepsilon_{j\pm 1}(0)$ and $d_{j\pm 1}$, from
Eq.~(\ref{2.38})
(together with the properties of the permittivity) it follows
that (\mbox{$\varepsilon_j$ $\!=$ $\!1$})
\begin{equation}
\label{2.65}
\frac{1-\sqrt{\varepsilon_{j\pm 1}(0)}}{1+\sqrt{\varepsilon_{j\pm 1}(0)}}
\le\frac{\kappa_j-\kappa_{j\pm 1}}{\kappa_j+\kappa_{j\pm 1}} \le 0.
\end{equation}
Thus, using Eq.~(\ref{B10}) (and the corresponding equation for
$r^s_{l-}$) and writing  
\begin{equation}
\label{2.66}
r_{l\pm}^s =
\frac{\displaystyle{\frac{\kappa_l-\kappa_{l\pm 1}}
{\kappa_l+\kappa_{l\pm 1}}}
+ e^{-2\kappa_{l\pm 1} d_{l\pm 1}}\,r_{(l\pm 1)\pm}^s}
{1 + \displaystyle{\frac{\kappa_l-\kappa_{l\pm 1}}
{\kappa_l+\kappa_{l\pm 1}}}
\,e^{-2\kappa_{l\pm 1}d_{l\pm 1}}\,r_{(l\pm 1)\pm}^s}\,,
\end{equation}
we can establish that for \mbox{$l$ $\!=$ $\!j$} the inequalities
\begin{multline}
\label{2.67}
|r_{j\pm}^s|
\ge \frac{1+\sqrt{\varepsilon_{j\pm 1}(0)}}
{2\sqrt{\varepsilon_{j\pm 1}(0)}}
\ \times\\ \times \  
\left|\left|
\frac{\kappa_j-\kappa_{j\pm 1}}{\kappa_j+\kappa_{j\pm 1}}\right|
- e^{-2\kappa_{j\pm 1}d_{j\pm 1}} \bigl|r_{(j\pm 1)\pm}^s\bigr| \right|
\end{multline}
and
\begin{multline}
\label{2.68}
|r_{j\pm}^s|
\le {\textstyle\frac{1}{2}}\left[1+\sqrt{\varepsilon_{j\pm 1}(0)}\right]
\ \times\\ \times \  
\left|\frac{\kappa_j-\kappa_{j\pm 1}}{\kappa_j+\kappa_{j\pm 1}}
+ e^{-2\kappa_{j\pm 1}d_{j\pm 1}}r_{(j\pm 1)\pm}^s \right|
\end{multline}
are valid. The inequality (\ref{2.67}) shows that if
\begin{equation}
\label{2.69}
\lim_{\kappa_j\to 0+}|r_{(j\pm 1)\pm}^s|\ne \lim_{\kappa_j\to 0+}
\left|\frac{\kappa_j-\kappa_{j\pm 1}}{\kappa_j+\kappa_{j\pm 1}}\right|
\end{equation}
holds, then (for the chosen values of $\phi$)
$r_{j\pm}^{s}$ approaches non-zero values 
in the limit \mbox{$\kappa_j$ $\!\to$ $\!0+$},
making Eq.~(\ref{2.61}) impossible. Hence,
the $d_{j}^{-4}$ law [Eq.~(\ref{2.55})] can be expected to hold
for large distances. By contrast, if
there are \emph{single-slab} (dielectric) 
walls, then Eq.~(\ref{2.66}) yields for
\mbox{$l$ $\!=$ $\!j$ $\!\pm$ $\!1$}
\begin{equation}
\label{2.70}
r_{(j\pm 1)\pm}^s
= \frac{\kappa_{j\pm 1}-\kappa_j}{\kappa_{j\pm 1}+\kappa_j}
\end{equation}
(\mbox{$r_{(j\pm 2)\pm}^s$ $\!=$ $\!0$}), and the inequality
(\ref{2.68}) thus implies that $r_{j\pm}^s$ vanishes
(uniformly with respect to $\phi$) as $\mathcal{O}(\kappa_{j} d_{j\pm1})$
in the limit \mbox{$\kappa_j$ $\!\to$ $\!0+$}, and a behavior as
in Eq.~(\ref{2.61}) is observed.
Note that from Eq.~(\ref{2.38}) for $\kappa_j$ and the
corresponding equation for $\kappa_{j+1}$ it follows
that the relation 
\begin{equation}
\label{2.71}
\kappa_{j+1}=\kappa_j
\sqrt{1+[\varepsilon_{j+1}(i\xi)-1]\cos^2 \phi}
\end{equation}
is valid, which reveals that $\kappa_{j+1}$ vanishes
with vanishing $\kappa_j$. Recall that according to Eq.~(\ref{2.50})
\mbox{$\kappa_j$ $\!\to$ $\!0+$} entails \mbox{$\xi$ $\!\to$
$\!0+$}. In a similar way it can be shown that (for single-slab walls) the
reflection coefficients for $p$-polarization,
$r_{j\pm}^p$, also vanish uniformly as $\mathcal{O}(\kappa_{j} d_{j\pm 1})$ in the
limit \mbox{$\kappa_j$ $\!\to$ $\!0+$}. Consequently, when the two walls are single-slab dielectrics,
than the $r_{j\pm}^\sigma$ behave according to Eq.~(\ref{2.61}),
and hence the $d_{j}^{-6}$-law is observed for large distances,
with the functions $R_{j\pm}^\sigma(\phi)$ in Eq.~(\ref{2.63}) being
proportional to the respective slab thickness.
Clearly, if only one of the two walls consists of a single slab, then
the $d_{j}^{-5}$-law may be observed. It should be pointed
out that the large-distance asymptotic regime again requires the
conditions (\ref{2.57}) and (\ref{2.60}) to be satisfied. 

Let us remark that it is conceivable that other
special choices of the walls may produce
other than $\sim\!d_{j}^{-4}$, $\sim\!d_{j}^{-5}$, and $\sim\!d_{j}^{-6}$
asymptotic distance dependences of the Casimir force.
It may also happen that in the asymptotic expansion of the
Casimir force terms $\sim\!d_{j}^{-n}$
with different values of $n$ must
be taken into account for not extremely large distances, if the
weights of the terms substantially
differ from each other. 
   
We finally note that the one-dimensional counterpart of the
$d_{j}^{-4}$-law is a $d_{j}^{-2}$-law.
It changes to a $d_{j}^{-4}$-law
when $r_{j\pm}\simeq \xi R_{j\pm}/c$
holds in the limit \mbox{$\xi$ $\!\to$ $\!0+$},
and it changes to a $d_{j}^{-3}$-law when only one of the
reflection coefficients shows this behavior.
In particular, two single-slab walls give rise
to a $d_{j}^{-4}$-law in place to the standard $d_{j}^{-2}$-law. 


\subsection{Short-distance law}
\label{sec3.3}

For short distances, we have to compare the $\kappa_{j}$-scale of
variation of the reflection coefficients $r_{j\pm}^\sigma$
with the now large scale of variation $d_{j}^{-1}$ of
$e^{-2\kappa_{j} d_{j}}$. In particular, assuming \mbox{$d_{j}$ $\!\ll d_{j \pm 1}$},
so that we may let
\begin{equation}
\label{2.72}
e^{-\kappa_{j\pm1} d_{j\pm1}} 
\approx e^{-\kappa_j d_{j\pm1}}
\approx e^{-d_{j\pm 1}/d_{j}} \approx 0,
\end{equation}
from Eqs.~(\ref{B10}) and (\ref{B11})
(and the corresponding equations for $r^\sigma_{l-}$) we see
that the reflection coefficients $r_{j\pm}^\sigma$
may be approximately replaced by
single-interface reflection coefficients according to
\begin{equation}
\label{2.73}
r^s_{j\pm}
\approx \frac{\kappa_j-\kappa_{j\pm 1}}{\kappa_j+\kappa_{j\pm 1}}\,,
\quad
r^p_{j\pm}
\approx \frac{\kappa_j/\kappa_{j \pm 1}-1/\varepsilon_{j\pm 1}}
{\kappa_j/\kappa_{j\pm 1} +1/\varepsilon_{j\pm 1}}\,.
\end{equation}
Thus, we effectively deal with two
semi-infinite walls of permittivities
$\varepsilon_{j\pm 1}$, so that Lifshitz' approximation for
short distances can be used
\cite{Lifshitz}:
\begin{equation}
\label{2.74}
\kappa_{j\pm 1} \approx \kappa_j \approx q
\end{equation}
and
\begin{equation}
\label{2.75}
r^s_{j\pm} \approx 0,
\quad
r^p_{j\pm}
\approx \frac{\varepsilon_{j\pm 1}-1}{\varepsilon_{j\pm 1}+1}\,.
\end{equation}
Note that Eqs.~(\ref{2.72}) and (\ref{2.74}) are consistent
with each other, and Eq.~(\ref{2.74}) is valid if  
\begin{equation}
\label{2.76}
\varepsilon_{j\pm 1}(\omega=i\kappa_j c)-1 \ll 1 ,
\end{equation}
i.e.
\begin{equation}
\label{2.77}
d_{j} \ll \frac{c}{\Omega_{j\pm 1}}\,,
\end{equation}
where the plasma frequencies $\Omega_{l}$
are defined by \mbox{$\Omega_l^2$ $\!=$
$\!\lim_{\omega\to\infty}(\omega^2\,[1$ $\!-$
$\!\varepsilon_l(\omega)])$} \cite{Jackson}.
In this approximation, Eq.~(\ref{2.43}) takes the
well-known form of \mbox{($v$ $\!=$ $\!2qd_{j}$)}
\begin{multline}
\label{2.78}
F\approx\frac{\hbar}{16\pi^2 d_{j}^3}  \int_0^\infty
d\xi \ \times\\
\times \ \int_0^\infty dv \,v^2
\left[\frac{(\varepsilon_{j+1} +1)}
{(\varepsilon_{j+1} -1)}\frac{(\varepsilon_{j-1}+1)}
{(\varepsilon_{j-1} -1)}e^v-1\right]^{-1}.
\end{multline}
Let us consider, for simplicity, single-resonance media
of Drude-Lorentz type, such that   
\begin{equation}
\label{2.79}
\varepsilon_{j\pm 1}(\omega) 
= 1 - \frac{\Omega^2}
{\omega^2+i\gamma_{0}\omega-\omega_{0}^2}\,.
\end{equation}
For small $\gamma_0$, Eq.~(\ref{2.78}) can then be further
evaluated to obtain (Appendix \ref{Evaluation})
\begin{equation}
\label{2.80}
F \approx\frac{\hbar}{2\pi d_{j}^3}
\,\sqrt{\omega_{0}^2 + \Omega^2/2}
\,\widetilde{\rm Li}_2\!\left[
\frac{\Omega^4}{64(\omega_0^2 + \Omega^2/2)^2}
\right],
\end{equation}
%
with
\begin{equation}
\label{tildedilog}
\widetilde{\rm
Li}_{2}(z)=\frac{1}{2}\sum_{m=1}^{\infty}\frac{\Gamma(4m-1)}
{[\Gamma(2m)]^2}\,\frac{z^m}{m^3}\,.
\end{equation}


\section{Numerical results}
\label{sec4}

\begin{figure}[t]
\includegraphics[width=6cm]{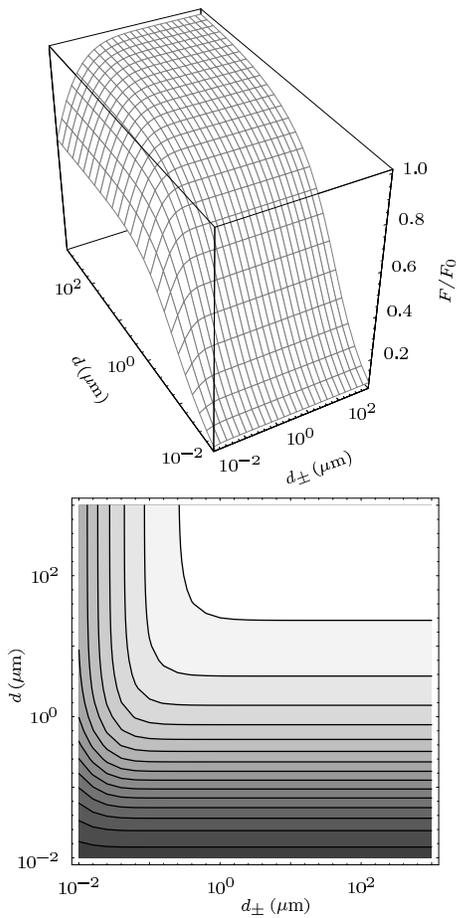}
\caption{\label{Mgdd}
Dependence of the relative Casimir force $F/F_0$ between two
identical single-slab walls on the thickness
$d_{\pm}$ of the walls and the wall separation $d$ (material parameters:  
$\omega_0=1.0 \times 10^{9}\, {\rm s}^{-1}=\omega_{\rm LO}$,
$\Omega=1.6176 \times 10^{16}\,{\rm s}^{-1}
\approx 10^8 \,\omega_{\rm LO}$,
$\gamma_0=9.7 \times 10^{14}\, {\rm s}^{-1}
\approx 10^6 \,\omega_{\rm LO}$).}
\end{figure}
\begin{figure}[t]
\includegraphics[width=6cm]{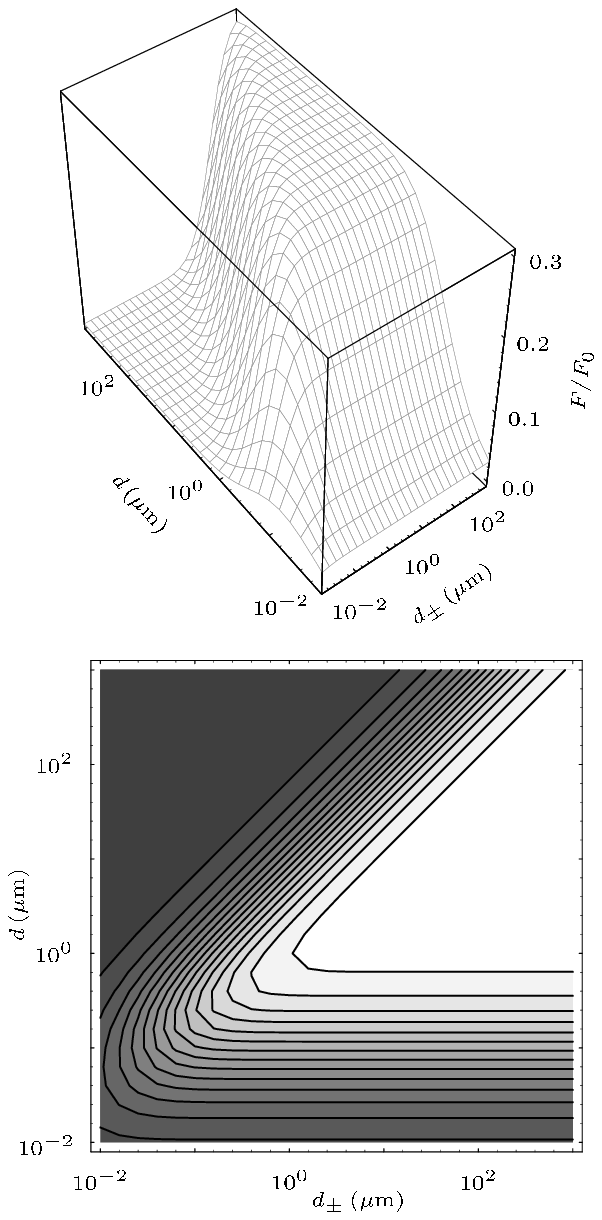}
\caption{\label{Sidd}
The same as in Fig.~\ref{Mgdd},
but with 
$\omega_{0}=2.0 \times 10^{15}\,{\rm s}^{-1}=\omega_{\rm HI}$,
$\Omega=6.536 \times 10^{15} \,{\rm s}^{-1}
\approx 3 \,\omega_{\rm HI}$, and
$\gamma_0=9.859 \times 10^{12} \,{\rm s}^{-1}
\approx  0.01 \,\omega_{\rm HI}$.}
\end{figure}
%
%
\begin{figure}[t]
\includegraphics[width=6cm]{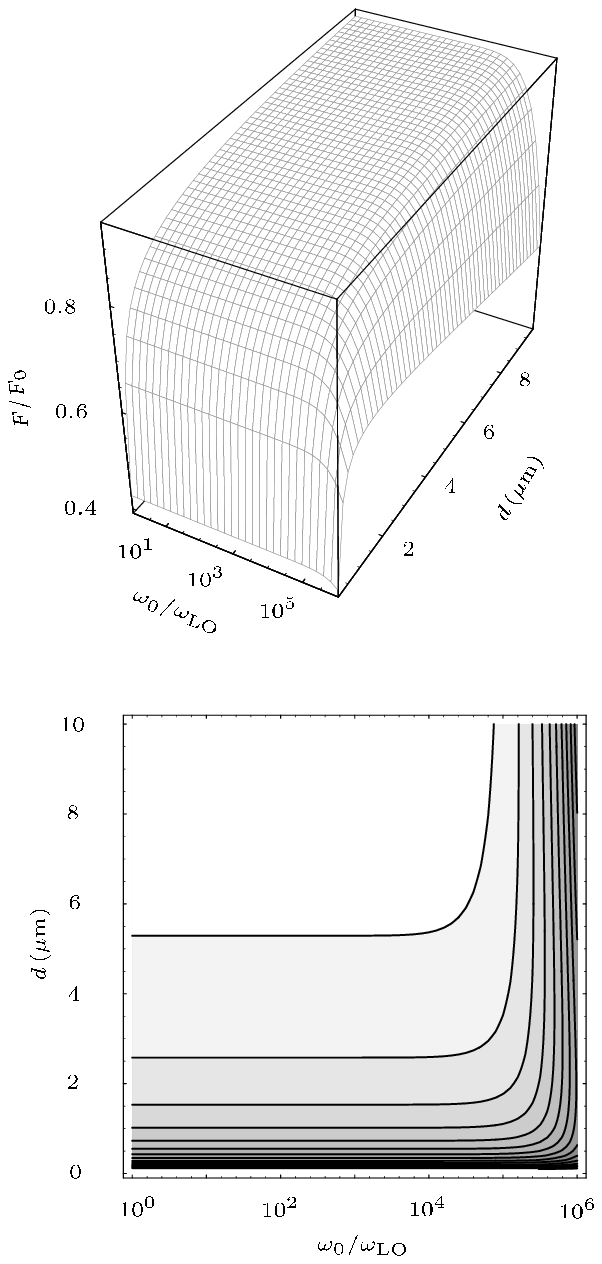}
\caption{\label{MgResonanz}
Dependence of the relative Casimir force $F/F_0$ between two
identical single-slab walls on the resonance frequency $\omega_0$
and the wall separation $d$ (wall thickness
$d_{\pm}=0.5\, \mu{\rm m}$; material parameters:
$\Omega=1.6176 \times 10^{16}\,{\rm s}^{-1}
\approx 10^8 \,\omega_{\rm LO}$,
$\gamma_0=9.7 \times 10^{14}\, {\rm s}^{-1}
\approx 10^6 \,\omega_{\rm LO}$).}
\end{figure}
\begin{figure}[t]
\includegraphics[width=6cm]{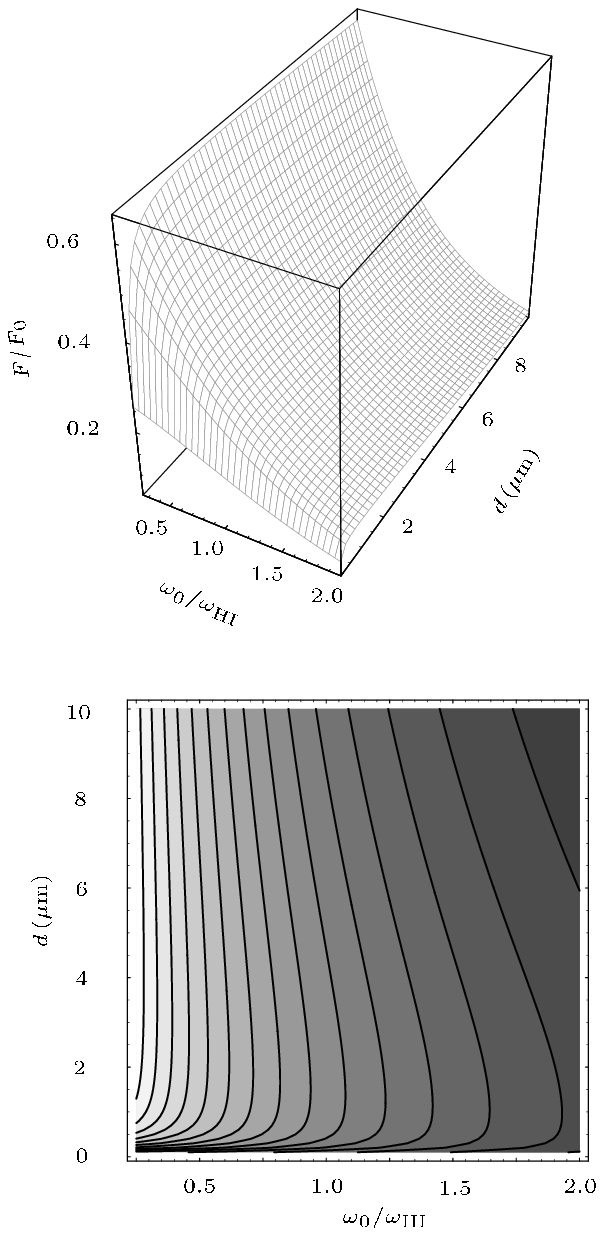}
\caption{\label{SiMnt}
The same as in Fig.~\ref{MgResonanz}, but with
$d_{\pm}=2.5\, \mu{\rm m}$,
$\Omega=6.536 \times 10^{15}\, {\rm s}^{-1}
\approx  3 \,\omega_{\rm HI}$, and
$\gamma_0=9.859 \times 10^{12}\, {\rm s}^{-1}
\approx 0.01 \,\omega_{\rm HI}$.}
\end{figure}
%
%
\begin{figure}[t]
\includegraphics[width=6cm]{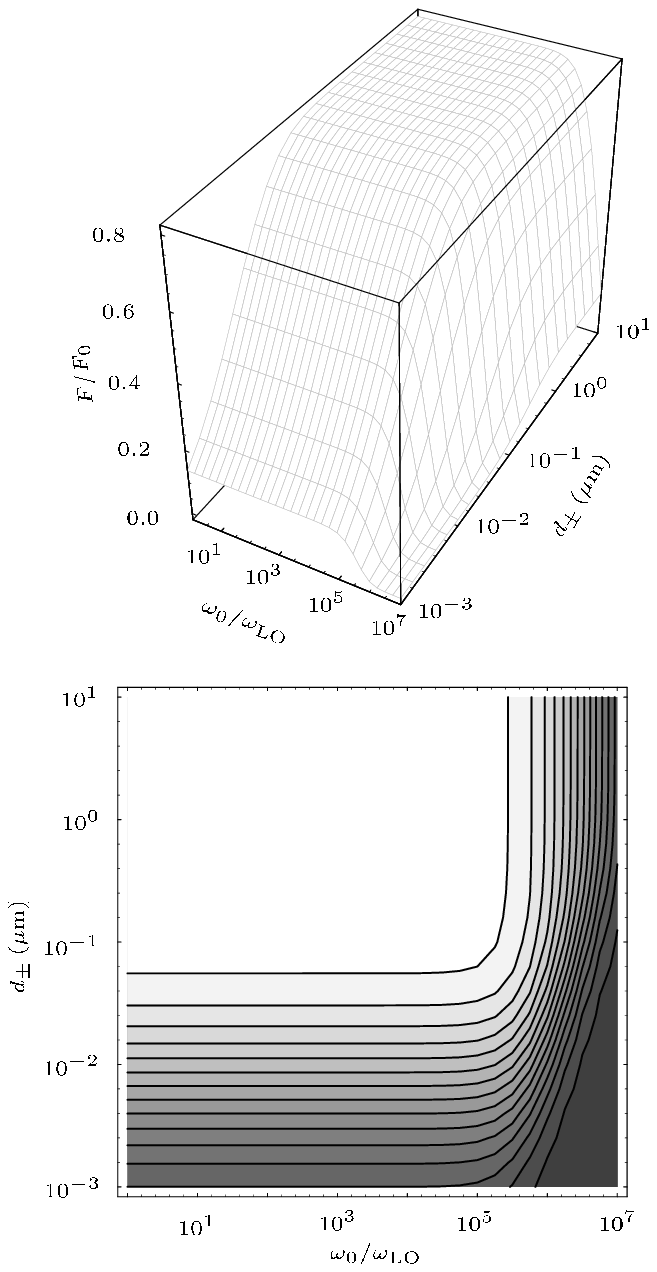}
\caption{\label{Mgd2Res}
Dependence of the relative Casimir force $F/F_0$ between two
identical single-slab walls on the resonance frequency $\omega_0$ and
the wall thickness $d_{\pm}$ (wall separation
$d=1\,\mu{\rm m}$; material parameters:
$\Omega=1.6176 \times 10^{16}\,{\rm s}^{-1}
\approx 10^8 \,\omega_{\rm LO}$,
$\gamma_0=9.7 \times 10^{14}\, {\rm s}^{-1}
\approx 10^6 \,\omega_{\rm LO}$).}
\end{figure}
\begin{figure}[t]
\includegraphics[width=6cm]{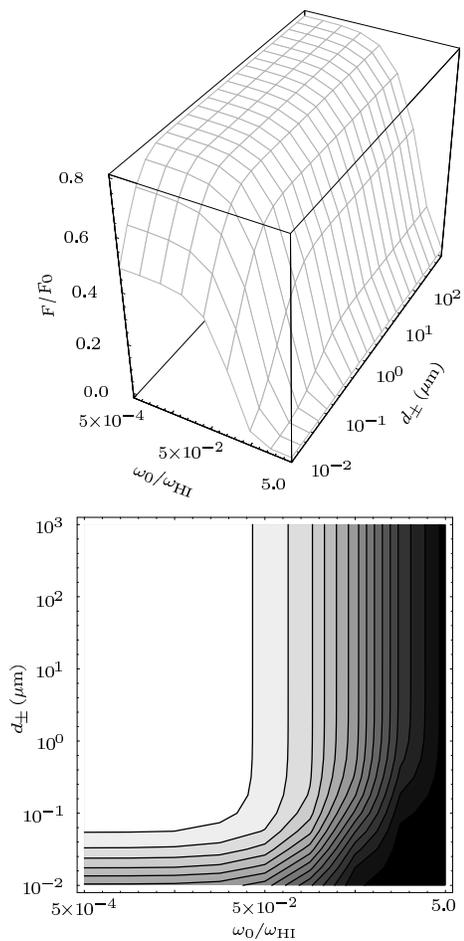}
\caption{\label{Sid2Res}
The same as in Fig. \ref{Mgd2Res}, but with
$d=1\,\mu{\rm m}$,
$\Omega=6.536 \times 10^{15}\, {\rm s}^{-1}
\approx  3 \,\omega_{\rm HI}$, and
$\gamma_0=9.859 \times 10^{12}\, {\rm s}^{-1}
\approx  0.01 \,\omega_{\rm HI}$.
The lowest contours are not trustworthy, because
of numerical errors.}
\end{figure}
%
%

\begin{figure}[t]
\includegraphics[width=6cm]{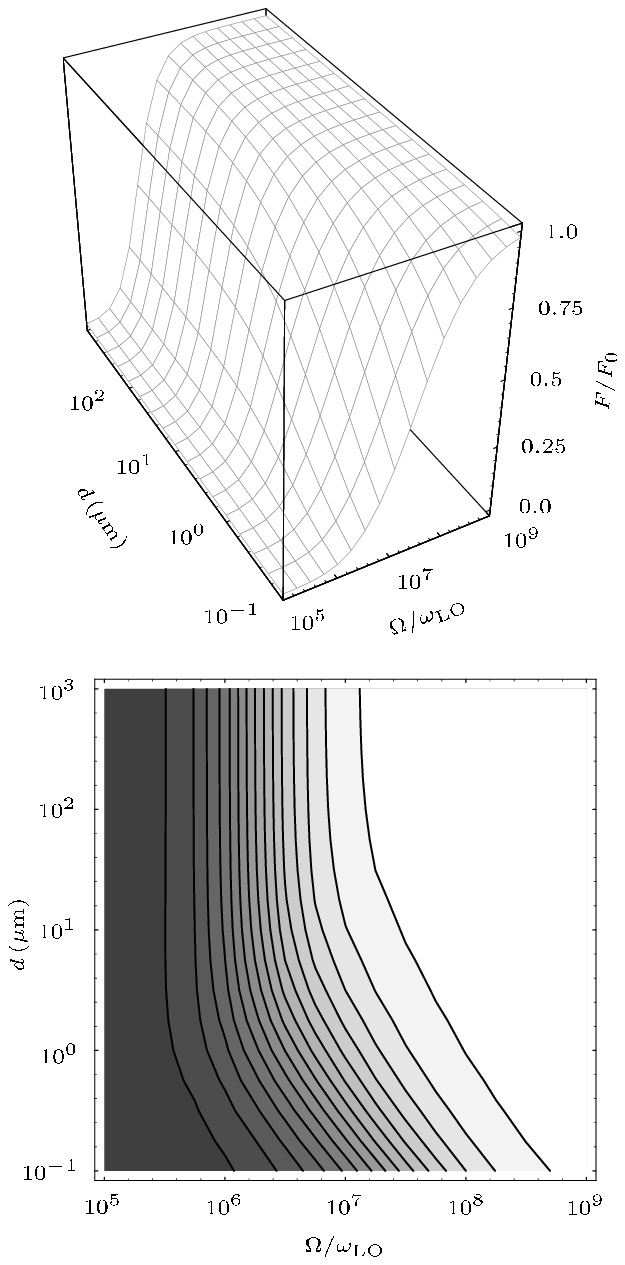}
\caption{\label{MgPlasma}
Dependence of the relative Casimir force $F/F_0$ between two
identical single-slab walls on the plasma frequency $\Omega$
and the wall separation $d$ (wall thickness
$d_{\pm}=0.5\,\mu{\rm m}$; material parameters:  
$\omega_0=1.0 \times 10^{9} \,{\rm s}^{-1}=\omega_{\rm LO}$,
$\gamma_0=9.7 \times 10^{14} \,{\rm s}^{-1}
\approx 10^6 \,\omega_{\rm LO}$).}
\end{figure}

\begin{figure}[t]
\includegraphics[width=6cm]{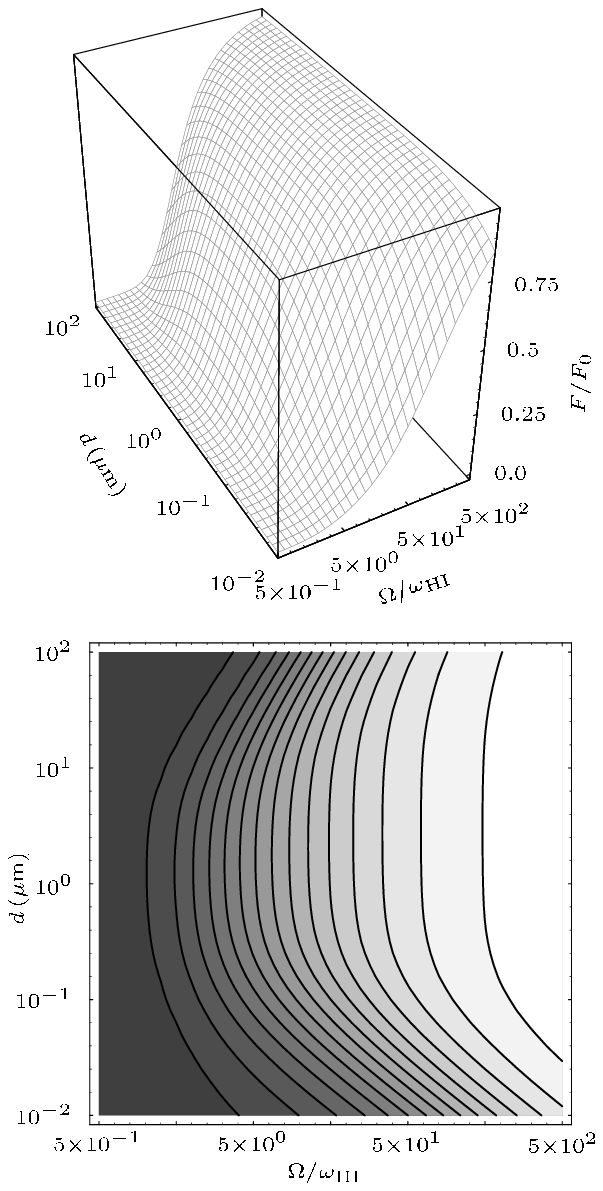}
\caption{\label{SiPlasma}The same as in Fig.~\ref{MgPlasma}, but with 
$d_{\pm}=2.5\,\mu{\rm m}$,   
$\omega_{0}=2.0 \times 10^{15}\,{\rm s}^{-1}=\omega_{\rm HI}$,
and $\gamma_0=9.859 \times 10^{12} \,{\rm s}^{-1}
\approx  0.01 \,\omega_{\rm HI}$.}
\end{figure}


%
\begin{figure}[t]
\includegraphics[width=6cm]{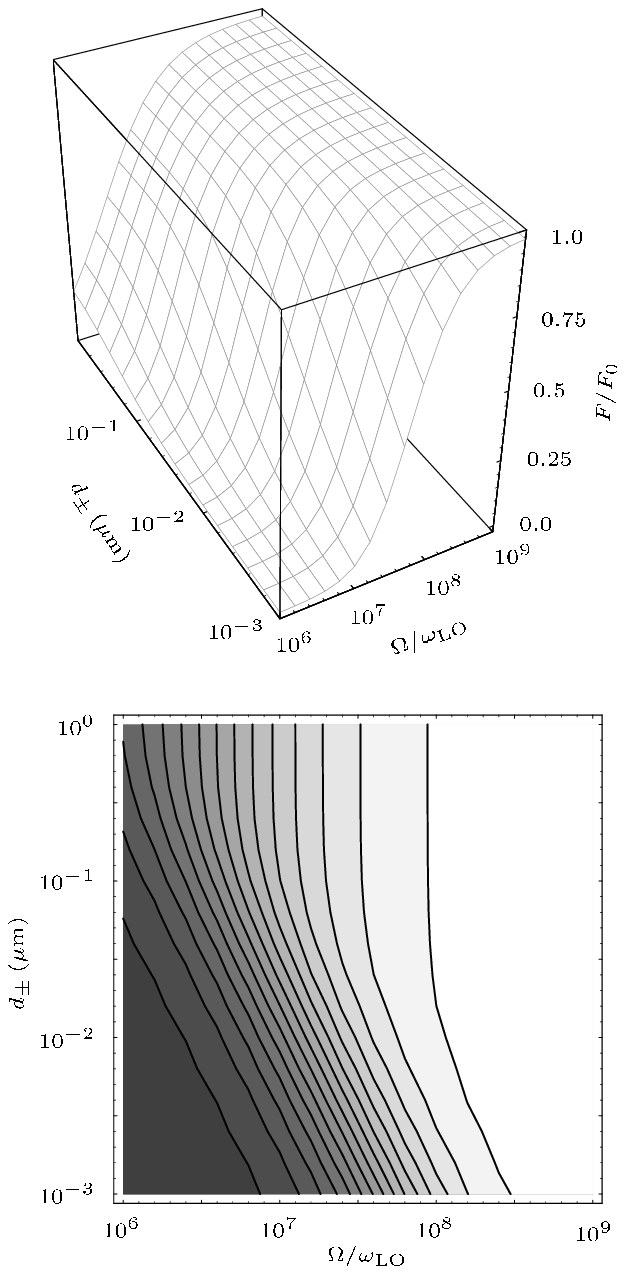}
\caption{\label{Mgd2Plasma}
Dependence of the relative Casimir force $F/F_0$ between two
identical single-slab walls on the plasma frequency $\Omega$
and the wall thickness $d_\pm$ (wall separation
$d=1\,\mu{\rm m}$; material parameters:  
$\omega_0=1.0 \times 10^{9} \,{\rm s}^{-1}=\omega_{\rm LO}$,
$\gamma_0=9.7 \times 10^{14} \,{\rm s}^{-1}\approx 10^6 \,\omega_{\rm LO}$).}
\end{figure}

\begin{figure}[t]
\includegraphics[width=6cm]{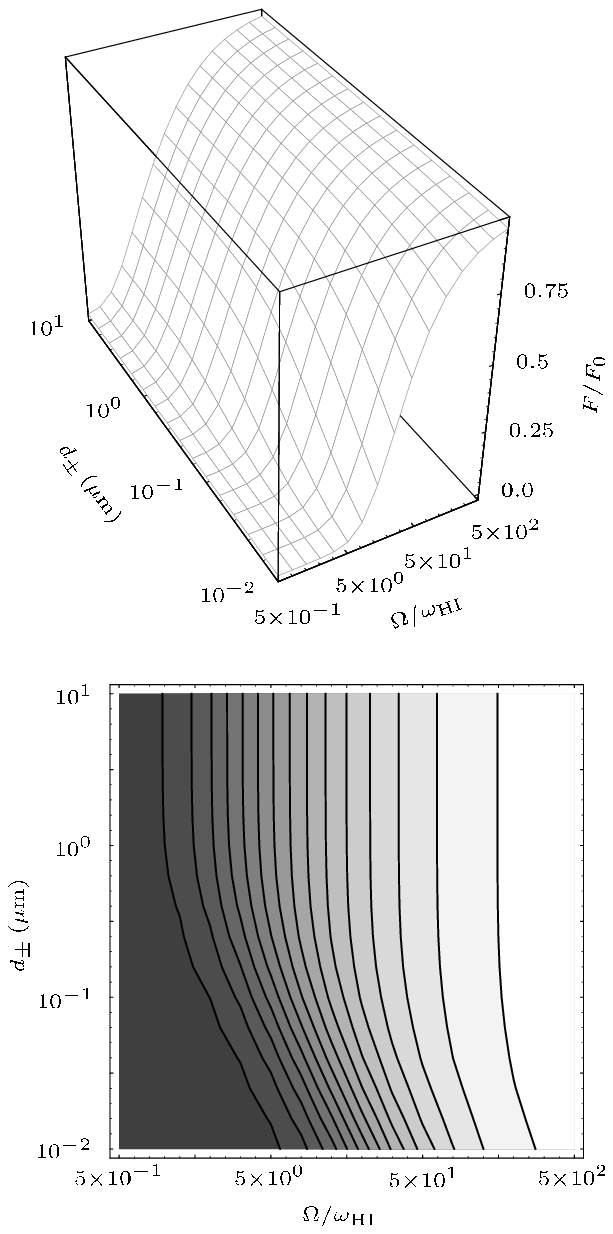}
\caption{\label{Sid2Plasma}
The same as in Fig.~\ref{Mgd2Plasma}, but with
$d=1\,\mu{\rm m}$,
$\omega_{0}=2.0 \times 10^{15}\,{\rm s}^{-1}=\omega_{\rm HI}$,
and $\gamma_0=9.859 \times 10^{12} \,{\rm s}^{-1}
\approx  0.01 \,\omega_{\rm HI}$.}
\end{figure}
%
%
\begin{figure}[t]
\includegraphics[width=6cm]{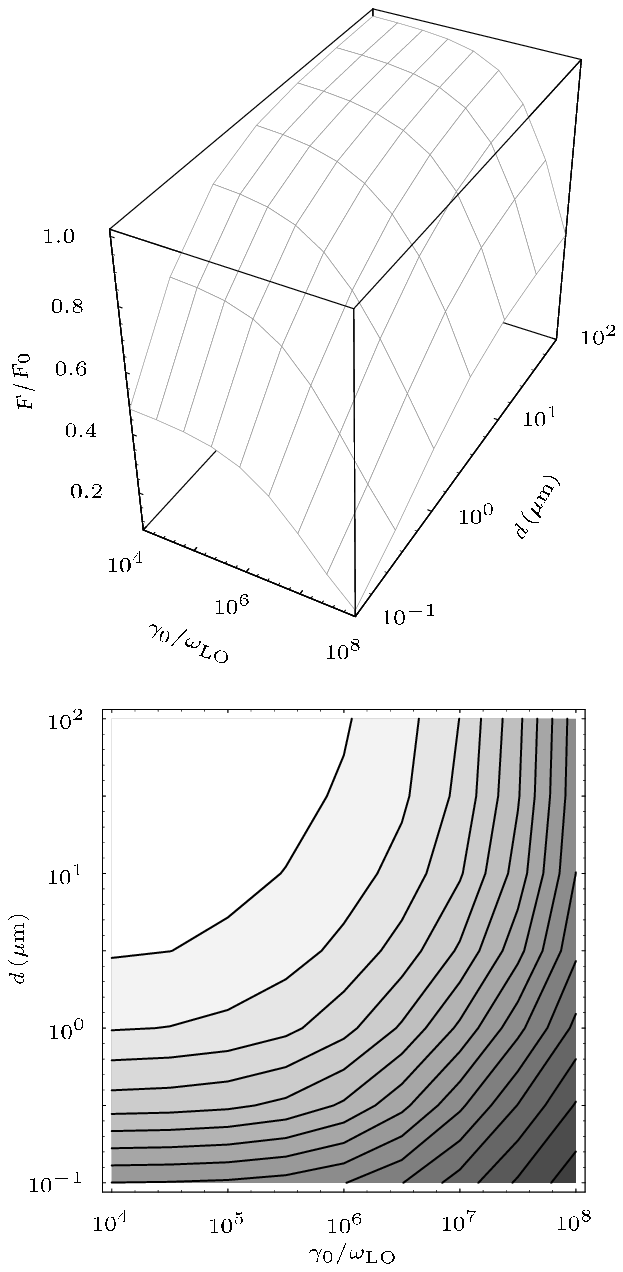}
\caption{\label{Mgdamp}
Dependence of the relative Casimir force $F/F_0$ between two
identical single-slab walls on the absorption parameter $\gamma_0$
and the wall separation $d$ (wall thickness
$d_{\pm}=0.5 \,\mu{\rm m}$; material parameters:  
$\omega_0=1.0 \times 10^{9} \,{\rm s}^{-1}=\omega_{\rm LO}$,
$\Omega=1.6176 \times 10^{16}\,{\rm s}^{-1}
\approx 10^8 \,\omega_{\rm LO}$).}
\end{figure}

\begin{figure}[t]
\includegraphics[width=6cm]{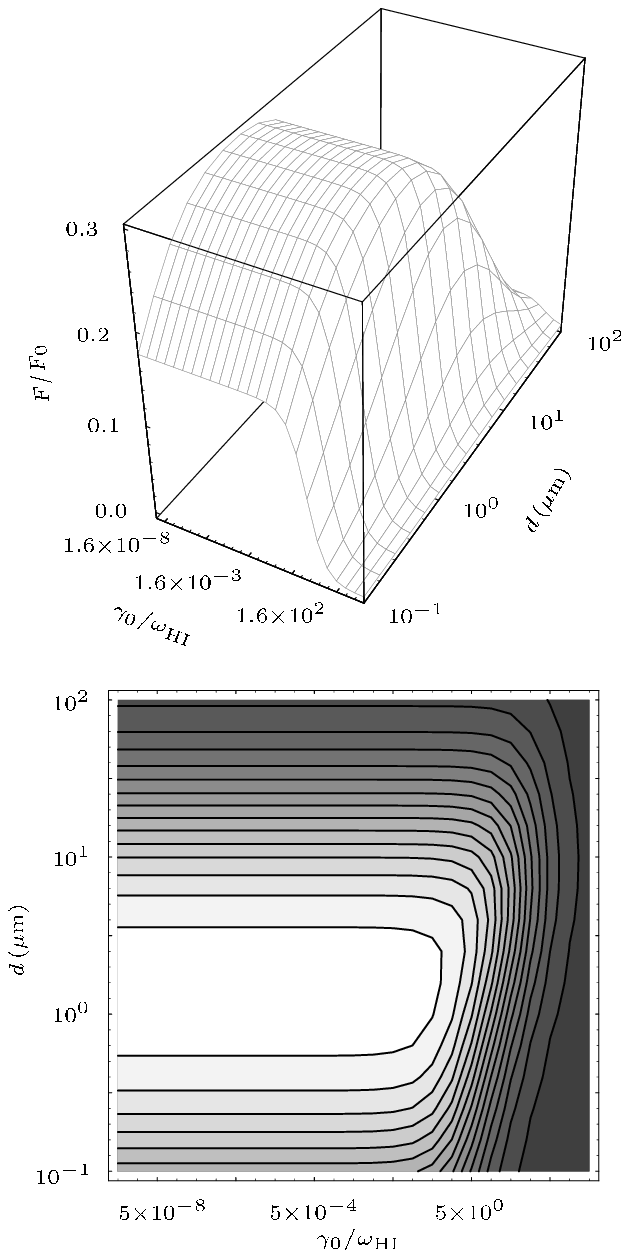}
\caption{\label{Sidamp}
The same as in Fig.~\ref{Mgdamp}, but with 
$d_{\pm}=2.5\,\mu{\rm m}$,  
$\omega_{0}=2.0 \times 10^{15}\,{\rm s}^{-1}=\omega_{\rm HI}$,
and $\Omega=6.536 \times 10^{15} \,{\rm s}^{-1}
\approx  3 \,\omega_{\rm HI}$.}
\end{figure}
%
%
\begin{figure}[t]
\includegraphics[width=6cm]{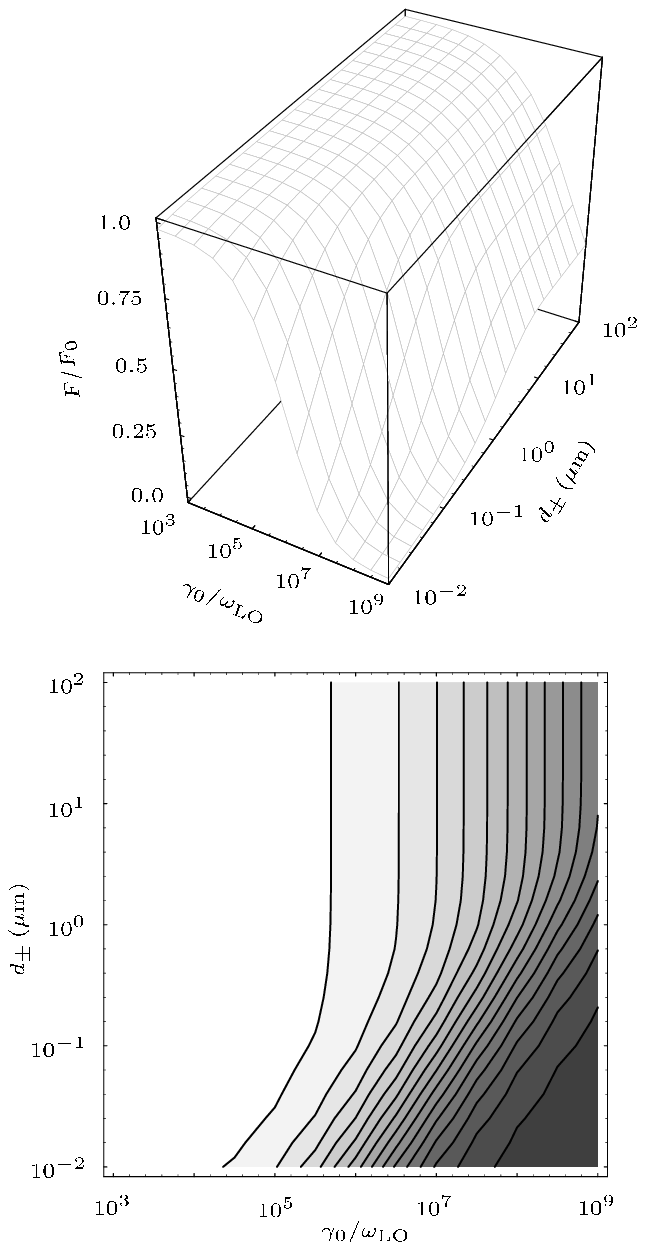}
\caption{\label{Mgd2damp}
Dependence of the relative Casimir force $F/F_0$ between two
identical single-slab walls on the absorption parameter $\gamma_0$
and the wall thickness $d_\pm$ (wall separation
$d=10\,\mu{\rm m}$; material parameters:  
$\omega_0=1.0 \times 10^{9} \,{\rm s}^{-1}=\omega_{\rm LO}$,
$\Omega=1.6176 \times 10^{16}\,{\rm s}^{-1}
\approx 10^8 \,\omega_{\rm LO}$).
}
\end{figure}

\begin{figure}[t]
\includegraphics[width=6cm]{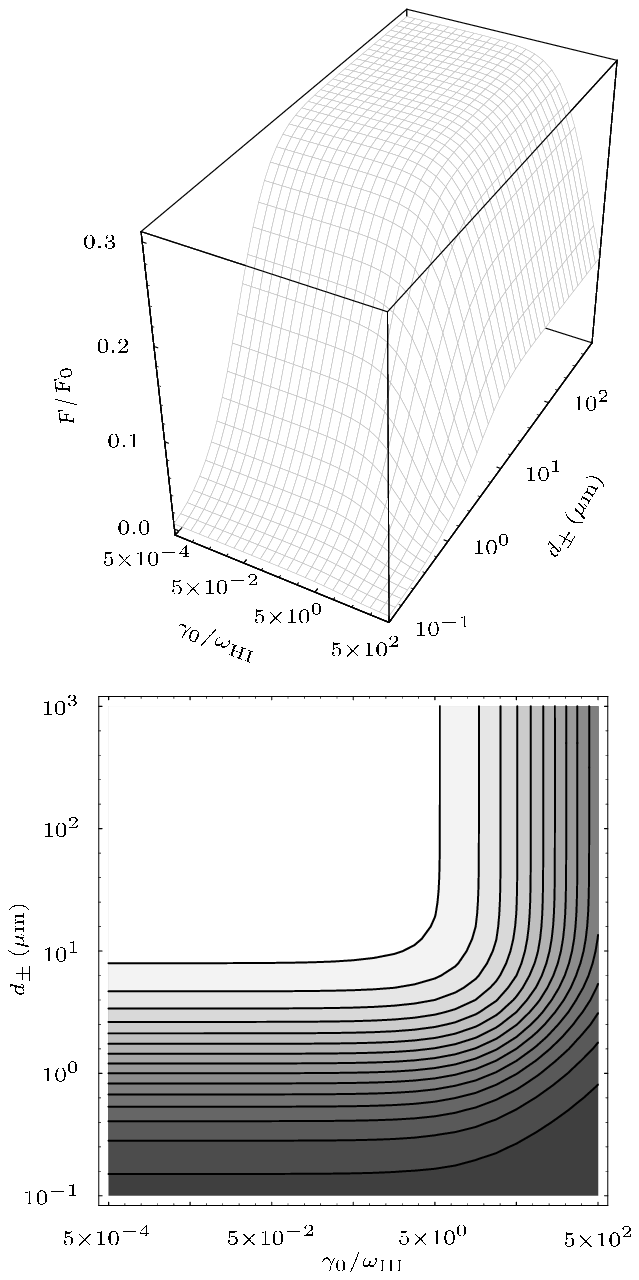}
\caption{\label{Sid2damp}
The same as in Fig.~\ref{Mgd2damp}, but with 
$d=10\,\mu{\rm m}$,  
$\omega_{0}=2.0 \times 10^{15}\,{\rm s}^{-1}=\omega_{\rm HI}$,
and $\Omega=6.536 \times 10^{15} \,{\rm s}^{-1}
\approx  3 \,\omega_{\rm HI}$.}
\end{figure}
\begin{figure}[t]
\includegraphics[width=6cm]{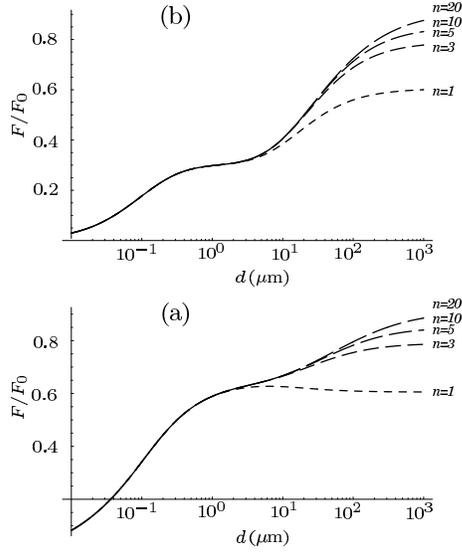}
\caption{\label{multi1to20}
Dependence of the relative Casimir force $F/F_0$ between two
identical multilayered walls on the wall separation $d$
for various numbers $n$ of identical bilayers (a) L-H and
(b) H-L each wall is
composed of. The data of the individual layers L and H are  
respectively
$d_{\pm} =0.01\,\mu{\rm m}$,
$\omega_{0}=\omega_{\rm LO} = 1.0 \times 10^{9}\,{\rm s}^{-1}$,
$\Omega=1.6176 \times 10^{16}\,{\rm s}^{-1}$,
$\gamma_0=9.7 \times 10^{14}\,{\rm s}^{-1}$
and
$d_{\pm} =2\,\mu{\rm m}$,
$\omega_{0}=\omega_{\rm HI} = 2.0 \times 10^{15}\,{\rm s}^{-1}$,
$\Omega=6.536 \times 10^{15}\,{\rm s}^{-1}$,
$\gamma_0=9.859 \times 10^{12}\,{\rm s}^{-1}$.}
\end{figure}
\begin{figure}[t]
\includegraphics[width=6cm]{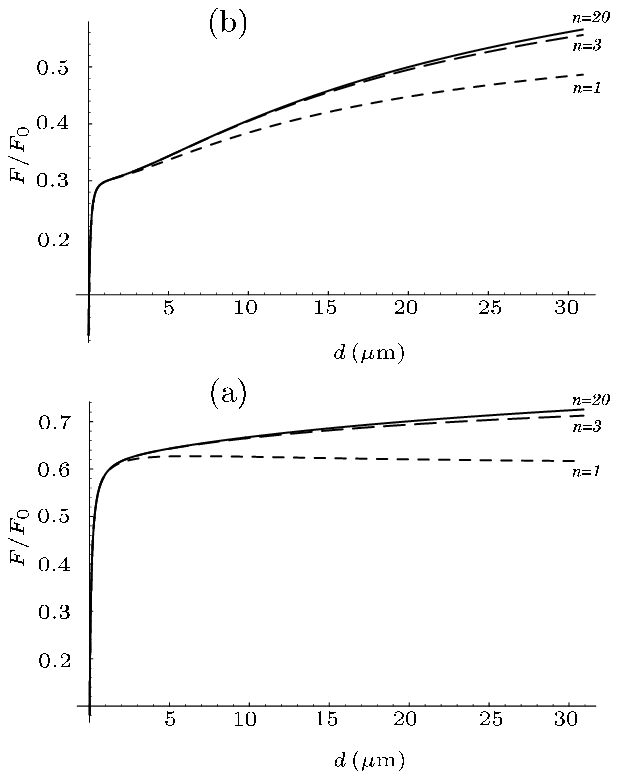}
\caption{\label{multi1320lin}
Detail of Fig.~\ref{multi1to20}.}
\end{figure}

%
In order to illustrate the dependence of
the Casimir force on the various parameters, we
have evaluated Eq.~(\ref{2.53}) numerically for
single-slab walls (Figs.~\ref{Mgdd} -- \ref{Sid2damp})
and a periodic multilayer wall structure (Figs.~\ref{multi1to20}
and \ref{multi1320lin}). The results extend, in a sense, the
1D results given in \cite{MexicanGuys} to three dimensions.
In Figs.~\ref{Mgdd} -- \ref{Sid2damp}, the wall material
is characterized by a single-resonance Drude-Lorentz permittivity
of the type given in Eq.~(\ref{2.79}).
We have performed the calculations for transverse
resonance frequencies $\omega_0$ in two qualitatively different
frequency domains, namely
\begin{equation}
\label{4.2}
\omega_0 \simeq \omega_{\rm LO}=1.0\times 10^{9}\,{\rm s}^{-1}
\end{equation}
and
\begin{equation}
\label{4.1}
\omega_0 \simeq \omega_{\rm HI}=2.0\times 10^{15}\,{\rm s}^{-1}.
\end{equation}
While the higher frequency $\omega_{\rm HI}$
corresponds to dielectric material (such as Si),
the lower frequency $\omega_{\rm LO}$ may be regarded
as being typical of metal-like material (such as Mg). 
To allow a comparison with the 1D results in Ref.~\cite{MexicanGuys},
we have performed the numerical calculations for the parameter values
used therein. Note that in the contour plots there are always $15$
lines of equal relative force $F/F_0$, equidistantly between
the highest and the lowest occurring value.

Let us first consider the relative force
$F/F_0$ between two single-slab walls.
The influence of the wall thickness 
\mbox{$d_\pm$ $\!\equiv\! d_{j\pm 1}$}
on the distance law for the two resonance
frequencies \mbox{$\omega_0$ $\!=\!\omega_{\rm LO}$}
and \mbox{$\omega_0$ $\!=\!\omega_{\rm HI}$}, respectively,
is shown in Figs.~\ref{Mgdd} and \ref{Sidd}
\mbox{($d$ $\!\equiv\! d_{j}$)}, and the
dependence on $\omega_0$ of $F/F_0$ for chosen $d_\pm$
is shown in Figs.~\ref{MgResonanz} and \ref{SiMnt}.
From Figs.~\ref{Sidd} and \ref{SiMnt} it is seen that
(for sufficiently high resonance frequencies)
the relative force increases with the distance between the
plates, attains a maximum, and eventually decreases with further
increasing plate separation. With increasing thickness of the plates
the maximum becomes broader and a plateau-like behavior
is observed. Figures \ref{Mgdd} and \ref{MgResonanz}
reveal that for low resonance frequencies the plateau can
become very broad, so that observation of decreasing values
of $F/F_0$ would require very large distances, at which
the force effectively vanishes. Comparing Figs.~\ref{Mgdd}
and \ref{Sidd}, we see that the response of the force to a change of
the plate thickness is much more sensitive for high
resonance frequencies than for low ones. Needless to say
that for sufficiently thick plates the force becomes
independent of the plate thickness. From Fig.~\ref{Sidd} it is seen
that the long-distance asymptotic behavior \mbox{$F$ $\!\sim$ $\!d^{-6}$}
[Section \ref{sec3.2}] is observed when -- in agreement with
the condition (\ref{2.60}) -- the distance between the plates
substantially exceeds the plate thickness.
A comparison of the 3D results in  Figs.~\ref{Mgdd}
and \ref{Sidd} with the 1D results in Fig.~2 in
Ref.~\cite{MexicanGuys} shows a quantitatively rather than
qualitatively different behavior of the relative force $F/F_0$
in the two theories. Clearly, the force itself behaves 
quite different in the two theories.     

Figures \ref{MgResonanz} and \ref{SiMnt} clearly show
that increasing the resonance frequency $\omega_0$ generally
lowers the force. In particular, it is seen that the position
of the maximum of the (relative) force is shifted to smaller
values of the wall separation when the resonance frequency
increases. At the same time, the maximum value decreases
and the long-distance asymptotic behavior sets in at smaller
distances. The dependence of the force on the resonance frequency
is in agreement with the Drude-Lorentz permittivity
(\ref{2.79}). For chosen plasma frequency $\Omega$ and
absorption parameter $\gamma_0$, the maximum absolute value of the
permittivity decreases with increasing value of $\omega_0$, thus
reducing the reflection coefficients of the walls.
From Figs.~\ref{Mgd2Res} and \ref{Sid2Res} it is seen
that increasing the value of the resonance frequency has a
similar effect as decreasing the value of the wall thickness. Both a
very small plate thickness and a low permittivity can lead to
poor plate reflectivity.

The influence of the plasma frequency
on the distance law for chosen plate thickness and the two
resonance frequencies \mbox{$\omega_0$ $\!=\!\omega_{\rm LO}$}
and \mbox{$\omega_0$ $\!=\!\omega_{\rm HI}$} is shown in
Figs.~\ref{MgPlasma} and \ref{SiPlasma}, respectively.
Figures \ref{Mgd2Plasma} and \ref{Sid2Plasma} illustrate
the dependence of the force on the plasma frequency and
the thickness of the plates for chosen distance between them.
Since the plasma frequency can be regarded as being a measure of
the strength of the medium resonance, higher values of the plasma frequency
imply higher values of the plate reflectivity and thus higher
values of the force, as can be clearly seen from
the figures. Moreover, the width of the
band-gap featured by the permittivity (\ref{2.79})
increases with the plasma frequency, which explains
that the interval of large (relative) force also grows with
increasing plasma frequency (Fig.~\ref{SiPlasma}).
In particular, Figs.~\ref{Mgd2Plasma} and \ref{Sid2Plasma} show
that a high plasma frequency can compensate for a small
plate thickness. Note that $\Omega\to\infty$ corresponds
to perfectly reflecting plates, whose thickness can then be
arbitrarily small.

The effect of material absorption is illustrated in
Figs.~\ref{Mgdamp} -- \ref{Sid2damp}. The dependence of the
distance law on the absorption parameter $\gamma_0$ for chosen plate
thickness and the two resonance frequencies \mbox{$\omega_0$ $\!=\!\omega_{\rm LO}$}
and \mbox{$\omega_0$ $\!=\!\omega_{\rm HI}$}, respectively,
is shown in Figs.~\ref{Mgdamp} and \ref{Sidamp}, and 
Figs.~\ref{Mgd2damp} and \ref{Sid2damp} present
the dependence of the force on the absorption parameter and
the thickness of the plates for chosen distance between them.
As expected, the force is seen to decrease with increasing
absorption parameter. The effect is similar to that observed
when the resonance frequency is increased. In both cases
the maximum absolute value of the
permittivity decreases (for chosen plasma frequency), so that the 
reflection coefficients of the walls diminish. It is worth
noting that the force responds more sensitively to a change
of $\gamma_0/\omega_0$ for high resonance frequencies than
for low ones. In particular, from Fig.~\ref{Sidamp}
it is seen that in the first case the force is
practically not influenced by material absorption as long
as \mbox{$\gamma_0/\omega_0$ $\!\ll$ $\!1$} is valid, and
it effectively reduces to zero when $\gamma_0/\omega_0$
substantially exceeds unity. Figure \ref{Sidamp} also shows
that the maximum of the relative force $F/F_0$ is shifted to
larger values of the distance between the plates when the value of
$\gamma_0/\omega_0$ increases. For low resonance frequencies
the ratio $\gamma_0/\omega_0$ must increase to rather
extreme values before the force substantially diminishes,
as it is seen from Fig.~\ref{Mgdamp}.
Note that for a metal ($\omega_0$ $\!\to$ $\!0$), $\gamma_0$
becomes inversely proportional to the conductivity. Small $\omega_0$
and small $\gamma_0$, i.e. high conductivity, lead to such a broad plateau of nearly constant (maximum)
value of $F/F_0$ $\!\approx$ $\!1$ that effectively
Casimir's formula (\ref{2.54}) applies.

Let us finally consider the force between two identical 
multilayered walls composed of identical bilayers, where
each bilayer is made of a metal-like and a dielectric-like
material (Figs.~\ref{multi1to20} and \ref{multi1320lin}).
For comparison with the 1D results in Ref.~\cite{MexicanGuys},
we have again performed the numerical calculations for the
parameter values used therein.
From Fig.~\ref{multi1to20} it is seen that
for sufficiently small distances between the walls the force does not
depend on the number of bilayers the walls are composed of.
In this case, only the inner layers essentially determine
the force, which is in full agreement with Eq.~(\ref{2.73}). 

For larger distances between the walls the relative
force increases with the number of bilayers.
The changes in the curvature of the curves in the figures
indicate that (for chosen number of bilayers) the distance law
can drastically change several times before the relative
forces becomes constant. These changes, which are less pronounced
when the inner layers are metal-like ones 
[Figs.~\ref{multi1to20}(a) and \ref{multi1320lin}(a)],
may be regarded as being typical of a 3D theory
(cf. Ref.~\cite{MexicanGuys}).


\section{Summary}
\label{summary}

We have studied the Casimir force between dispersing and absorbing
multilayered dielectric plates. On the basis of the quantization
scheme for the electromagnetic field in causal media as given
in Ref.~\cite{Welsch} we have extended the recently derived
zero-temperature result \cite{TomasCasimir} to finite temperatures.
The derived formula generalizes Lifshitz' formula \cite{Lifshitz} to
arbitrary multilayered walls, and application of the
1D version of the theory to single-slab walls yields the
results in Ref.~\cite{Kup}.

Restricting our attention to the zero-temperature li\-mit, we
have studied the problem of asymptotic distance laws of the
Casimir force. We have
shown that Lifshitz' approximation for short distances
also applies to multilayered walls and, depending on the
wall structure, the distance law can
drastically change with increasing wall separation. In particular,
for two single-slab walls the Casimir force
tends to behave like $d^{-6}$ instead of $d^{-4}$ as the wall
separation $d$ goes to infinity. 

Assuming permittivities of Drude-Lorentz type, we
have finally presented a number of numerical results in order to
illustrate the dependence of the zero-temperature Casimir force on
various system parameters and to compare with the 1D results
recently reported in Ref.~\cite{MexicanGuys}. While
in the case of single-slab walls the 3D and the 1D theory yield
a qualitatively similar dependence of the relative Casimir force 
on the wall separation, significant differences (not only for
the absolute force but even for the relative force) may be observed for
more complicated, multilayered
walls.
%
\begin{acknowledgments}
C.R. would like to thank Ho Trung Dung
for useful advice concerning \scshape{Fortran}.
\end{acknowledgments}
\begin{appendix}

\section{Derivation of Eqs.~(\ref{2.22}), (\ref{2.23})}
\label{calculation}
It is convenient to discretize the
fundamental field variables $\hat{\bf f}({\bf r},\omega)$ so as to
work with $\hat{H} = \sum_{\mu,{\bf R}} \hbar\omega_\mu\,
\hat{\bf f}_{{\bf R},\mu}^\dagger \hat{\bf f}_{{\bf R},\mu}$ instead of
Eqs.~(\ref{2.20}) [$(\mathbf{R},\omega_\mu)$, grid points], and hence
with a partition function of the form
$Z = \prod_{\mu,{\bf R}} Z^3_{{\bf R},\mu}$.
From the continuum limit of such a calculation, it then
easily follows that
\begin{multline}
\label{A7}
\bigl\langle\hat{\mathbf{f}}^\dagger(\mathbf{r},\omega)\otimes
\hat{\mathbf{f}}(\mathbf{r'},\omega')\bigr\rangle
\\
=\tensor{\delta}(\mathbf{r,r'})\delta(\omega-\omega')
\,\frac{1}{2} \left[
\coth\!\left(\frac{{\hbar\omega}}{2k_{\rm B} T}\right) - 1
\right],
\end{multline}
\begin{multline}
\label{A8}
\bigl\langle\hat{\mathbf{f}}(\mathbf{r},\omega)\otimes
\hat{\mathbf{f}}^\dagger(\mathbf{r'},\omega')\bigr\rangle
\\
=\tensor{\delta}(\mathbf{r,r'})\delta(\omega-\omega')
\,\frac{1}{2}\left[
\coth\!\left(\frac{{\hbar\omega}}{2k_{\rm B} T}\right) + 1
\right]
\end{multline}
as well as
\begin{equation}
\label{A9}
\bigl\langle\hat{\mathbf{f}}(\mathbf{r},\omega)
\otimes \hat{\mathbf{f}}(\mathbf{r'},\omega')\bigr\rangle
=\bigl\langle\hat{\mathbf{f}}^\dagger(\mathbf{r},\omega)
\otimes \hat{\mathbf{f}}^\dagger(\mathbf{r'},\omega')\bigr\rangle = 0.
\end{equation}
Substituting Eq.~(\ref{2.0}) and the corresponding equation for
the displacement field into Eq.~(\ref{2.18}), we may write, on
recalling Eqs.~(\ref{2.11}), (\ref{2.12}), (\ref{2.13a}) and (\ref{A9})
\begin{eqnarray}
\label{A12}
\lefteqn{
\tensor{T}_1({\bf r},{\bf r}',t)
}
\nonumber\\&&
= \int_0^\infty d\omega \int_0^\infty d\omega'
\left[
e^{-i(\omega-\omega')t}
\bigl\langle\underline{\hat{\bf D}}({\bf r},\omega)
\otimes\underline{\hat{\bf E}}{^\dagger}({\bf r},\omega)
\bigr\rangle
\right.
\nonumber\\&&\hspace{10ex}
\left.
+\,e^{i(\omega-\omega')t}
\bigl\langle\underline{\hat{\bf D}}{^\dagger}({\bf r},\omega)
\otimes\underline{\hat{\bf E}}({\bf r},\omega)
\bigr\rangle 
\right].
\end{eqnarray}
Making here explicitly use of Eqs.~(\ref{2.11}) and (\ref{2.12}) and
the reciprocity property (\ref{2.13b}), we obtain
\begin{multline}
\label{A13}
\tensor{T}_1({\bf r},{\bf r}',t)\\
= \mu_0 \int_0^\infty d\omega \int_0^\infty d\omega' \,\omega'{^2}
\Nabla\times\Nabla\times
\tensor{K}(\mathbf{r,r'},\omega,\omega',t),
\end{multline}
where the tensor-valued function
$\tensor{K}(\mathbf{r,r'},\omega,\omega',t)$ is given by
\begin{multline}
\label{A14}
\tensor{K}(\mathbf{r,r'},\omega,\omega',t)
\\
= e^{-i(\omega-\omega')t}
\bigl\langle\bigl(\tensor{G} \star W{\bf f}\bigr)({\bf r},\omega)
\otimes\bigl(W{\bf f}^\dagger \star \tensor{G}^\ast\bigr)({\bf r}',\omega')
\bigr\rangle
\\
+ \,e^{i(\omega-\omega')t}
\bigl\langle\bigl(\tensor{G}^\ast \star W{\bf f}^\dagger\bigr)({\bf r},\omega)
\otimes\bigl(W{\bf f} \star \tensor{G}\bigr)({\bf r}',\omega')
\bigr\rangle.
\end{multline}
Here, the abbreviating notations
\begin{equation}
\label{A10}
W(\mathbf{r},\omega) \equiv
\sqrt{\hbar \varepsilon_0 \varepsilon''(\mathbf{r},\omega)/ \pi}
\end{equation}
and
\begin{equation}
\label{A11}
(\tensor{G} \star \hat{\mathbf{f}})(\mathbf{r},\omega)
\equiv \int d^3r' \,
\tensor{G}(\mathbf{r},\mathbf{r}',\omega)
\hat{\mathbf{f}}(\mathbf{r}',\omega)
\end{equation} 
have been used. Applying the integral relation (\ref{2.13c})
and using Eqs.~(\ref{A7}) and (\ref{A8}), from Eqs.~(\ref{A13})
and (\ref{A14}) we derive
\begin{multline}
\label{A15}
\tensor{T}_1({\bf r},{\bf r}')
\\
= \frac{\hbar}{\pi} \int_0^\infty d\omega\,
\coth\!\left(\frac{\hbar\omega}{2k_{\rm B}T}\right)
\Nabla\times\Nabla\times \Im[\tensor{G}({\bf r},{\bf r}',\omega)].
\end{multline}
Recalling Eq.~(\ref{2.13}), we see that Eq.~(\ref{A15}) just
leads to Eq.~(\ref{2.22}).
The magnetic part $\tensor{T}_2({\bf r},{\bf r}',t)$ [Eq.~(\ref{2.19})]
can be calculated analogously. In place of Eq.~(\ref{A12}) we now
have
\begin{eqnarray}
\label{A16}
\lefteqn{
\tensor{T}_2({\bf r},{\bf r}',t)
}
\nonumber\\&&
= \int_0^\infty d\omega \int_0^\infty d\omega'
\left[
e^{-i(\omega-\omega')t}
\bigl\langle\underline{\hat{\bf B}}({\bf r},\omega)
\otimes\underline{\hat{\bf H}}{^\dagger}({\bf r},\omega)
\bigr\rangle
\right.
\nonumber\\&&\hspace{10ex}
\left.
+\,e^{i(\omega-\omega')t}
\bigl\langle\underline{\hat{\bf B}}{^\dagger}({\bf r},\omega)
\otimes\underline{\hat{\bf H}}({\bf r},\omega)
\bigr\rangle 
\right],
\end{eqnarray}
from which by means of Eqs.~(\ref{2.10}) and
(\ref{2.12}) follows that 
\begin{multline}
\label{A17}
\tensor{T}_2({\bf r},{\bf r}',t)\\
= - \mu_0 \int_0^\infty d\omega \int_0^\infty d\omega'
\,\omega \omega'
\Nabla\times \tensor{K}(\mathbf{r,r'},\omega,\omega',t)
\times\Lnabla{'}.
\end{multline}
Comparing Eq.~(\ref{A17}) with Eqs.~(\ref{A13}) and (\ref{A15}),
we see that
\begin{multline}
\label{A18}
\tensor{T}_2({\bf r},{\bf r}')
\\
=-\frac{\hbar}{\pi}\int_0^\infty \!\!d\omega
\,\coth\! \left(\frac{\hbar\omega}{2k_{\rm B} T}\right)
\Nabla\times
\Im\!\left[\tensor{G}(\mathbf{r,r',\omega})\right]
\times\!\Lnabla{'}
\end{multline}
which is just Eq.~(\ref{2.23}).



\section{Reflection coefficients}
\label{Fresnel}

Let $\,\,\roarrow{\!\!E}_l^\sigma(z)$
be the classical (complex) electric field that is observed
in the $l$th layer when 
in a layer $l'$ with \mbox{$l'$ $\!<$ $\!l$}
an electromagnetic wave (of chosen frequency $\omega$, transverse
wave-vector component ${\mathbf q}$ and polarization $\sigma$)
propagates in $z$-direction
(i.e. upwards in Fig.~\ref{multilayer}). It can be
written as

\begin{equation}
\label{B0}
\,\roarrow{\!\!E}_l^\sigma(z)
= E_{+l'}^\sigma {t}_{l'/l}^\sigma\left[
{\bf e}^+_{\sigma l} e^{i\beta_l z}
+ e^{2i\beta_l d_l} r_{l+}^\sigma
{\bf e}^-_{\sigma l} e^{-i\beta_lz}\right],
\end{equation}
where $E_{+l'}^\sigma$ is the amplitude of the
upwards propagating wave just before the upper boundary
of the $l'$th layer, ${t}_{l'/l}^\sigma$ is the transmission
coefficient from the $l'$th to the $l$th layer \mbox{($l'$ $\!<$ $\!l$)},
and $r_{l+}^\sigma$ is the reflection coefficient at the
upper boundary of the $l$th layer (${\bf e}^\pm_{\sigma l}$ according
to Eq.~(\ref{2.29a}), \mbox{$\sigma$ $\!=$ $\!s,p$}).
Note that we adopt the convention \cite{Tomas} that $z=0$ denotes the
lower boundary in all layers $l$ with $l>0$, and $z=d_l$ denotes the upper boundary
in all layers $l$ with $l<n$. The transmission and reflection coefficients 
are determined by the requirement that
${\bf e}_t \,\roarrow{\!\!E}_{l}^\sigma(z)$  and
$({\bf e}_t \times \Nabla)\,\roarrow{\!\!E}_{l}^\sigma(z)$
are continuous at the surfaces of discontinuity
(${\bf e}_t$, tangential unit vector). Applying $\Nabla$ in the
$(\mathbf{q},z)$-space as $i\mathbf{q}+\mathbf{e}_{z}\partial/\partial
z$, straightforward calculation yields the sets of
equations
\begin{multline}
\label{B2}
{t}^s_{l'/l} e^{i\beta_l d_l } \bigl[1+r^s_{l+}\bigr]
\\
= {t}^s_{l'/l+1} \bigl[ 1+e^{ 2i \beta_{l+1}  d_{l+1} }
r^s_{(l+1)+} \bigr],
\end{multline}
\begin{multline}
\label{B4}
\beta_l {t}^s_{l'/l}
e^{i\beta_l d_l }\bigl[1-r^s_{l+}\bigr]
\\
= \beta_{l+1} {t}^s_{l'/l+1}
\bigl[1-e^{2i\beta_{l+1} d_{l+1}} r^s_{(l+1)+}\bigr]
\end{multline}
[\mbox{$k_l$ $\!=$ $\!\omega\varepsilon_l(\omega)^{1/2}/c$
$\!=$ $\!(q^2$ $\!+$ $\!\beta_l^2)^{1/2}$}] and

\begin{multline}
\label{B6}
\frac{\beta_l}{k_l}t^p_{l'/l}e^{i\beta_l d_l}
\bigl[1-r^p_{l+}\bigr]
\\
=\frac{\beta_{l+1}}{k_{l+1}}\,
{t}^p_{l'/l+1}\bigl[1-e^{2i\beta_{l+1}
d_{l+1} }r^p_{(l+1)+}\bigr],
\end{multline}
\begin{multline}
\label{B8}
k_l {t}^p_{l'/l} e^{ i\beta_l d_l }
\bigl[1+r^p_{l+}\bigr]
\\
= k_{l+1} {t}^p_{l'/l+1}\bigl[1+e^{2i\beta_{l+1} d_{l+1}}
r^p_{(l+1)+}\bigr]
\end{multline}
for $s$ and $p$ polarization, respectively. Eliminating in Eqs.~(\ref{B2} ) -- (\ref{B8})
${t}^\sigma_{l'/l}e^{i\beta_ld_l}/{t}^\sigma_{l'/l+1}$,
we derive the recurrences
\begin{equation}
\label{B10}
r^s_{l+}
= \frac{(\beta_l/\beta_{l+1}-1)
+(\beta_l/\beta_{l+1}+1)e^{2i\beta_{l+1} d_{l+1}} r^s_{(l+1)+}}
{(\beta_l/\beta_{l+1}+1)+(\beta_l/\beta_{l+1}-1)
e^{2i\beta_{l+1} d_{l+1}} r^s_{(l+1)+}}
\end{equation}
and
\begin{equation}
\label{B11}
r^p_{l+}
=\frac{(\frac{\beta_l}{\beta_{l+1}}
-\frac{\varepsilon_l}{\varepsilon_{l+1}})
+(\frac{\beta_l}{\beta_{l+1}}
+\frac{\varepsilon_l}{\varepsilon_{l+1}})
e^{2i\beta_{l+1} d_{l+1}} r^p_{(l+1)+}}
{(\frac{\beta_l}{\beta_{l+1}}+\frac{\varepsilon_l}{\varepsilon_{l+1}})
+(\frac{\beta_l}{\beta_{l+1}}
-\frac{\varepsilon_l}{\varepsilon_{l+1}})
e^{2i\beta_{l+1} d_{l+1} } r^p_{(l+1)+}},
\end{equation}
which terminate at \mbox{$l$ $\!=$ $\!n$ $\!-$ $\!1$}, because
of \mbox{$r^\sigma_{n+}$ $\!=$ $\!0$}.
Note that the $r^\sigma_{(n-1)+}$ for the last
surface of discontinuity are the
well-known single-interface coefficients.

By considering in the same
way the solution that emerges in the $l$th layer from a downwards
propagating wave in the $l'$th layer ($l'>l$), analogous recurrences
for the coefficients $r^\sigma_{l-}$ are derived. They are formally
obtained from Eqs.~(\ref{B10}) and (\ref{B11}) by making
the replacements
\begin{equation}
\label{B12}
l \mapsto l, \quad l+1 \mapsto l-1
\end{equation} 
and turning the subscript $+$ of the reflection coefficients into
$-$. Because of \mbox{$r_{0-}^\sigma$ $\!=$ $\!0$}, they
terminate at \mbox{$l$ $\!=$ $\!1$}.
From the recurrences
it follows that the reflection coefficients are real on the (positive!)
imaginary frequency axis and fulfill there
the inequality
\begin{equation}
-1\le r_{l\pm}^\sigma \le 1
\qquad
(\omega=i\xi).
\end{equation}
It should be noted that it is possible to give the $t^\sigma_{l'/l}$
explicitly once all the $r^\sigma_{l\pm}$ have been computed.


\section{Integration along the imaginary frequency axis}
\label{ContourFlip}
From Eq.~(\ref{2.13}) the
(exact) Lippmann-Schwinger-type integral equation
\begin{multline}
\label{C7}
\tensor{G}^{\rm scat}(\mathbf{r},\mathbf{r}',\omega)
\\
=\frac{\omega^2}{c^2}\int d^3s
\,\tensor{G}^{\rm bulk}(\mathbf{r},\mathbf{s},\omega)
\delta\varepsilon(\mathbf{s},\omega)
\tensor{G}(\mathbf{s},\mathbf{r}',\omega),
\end{multline}
can be derived,
where $\tensor{G}^{\rm bulk}$ and $\tensor{G}^{\rm scat}$ are the bulk and
scattering parts of the full Green tensor $\tensor{G}$, respectively,
and $\delta\varepsilon(\mathbf{r,\omega})$ is the associated deviation of the
permittivity from the `bulk' situation.
Since any possible Green tensor (such as $\tensor{G}$ or
$\tensor{G}^{\rm bulk}$) has an asymptotic behavior
$\mathcal{O}(c^2/\omega^2)$ for $|\omega|\to \infty$ in the upper
complex half-plane \cite{Welsch}, Eq.~(\ref{C7}) reveals
that $\omega^2\varepsilon({\bf r},\omega)
\tensor{G}^{\rm scat}({\bf r},{\bf r}',\omega)/c^2$
has exactly the same asymptotic
behavior as the permittivity difference
$\delta\varepsilon(\mathbf{r,\omega})$. We can therefore
conclude that frequency integrals of type
\begin{equation}
\label{C8}
I = \int_{\mathcal{C}} d\omega\,
\frac{\omega^2}{c^2}\varepsilon(\mathbf{r,\omega})
\tensor{G}^{\rm scat}(\mathbf{r,r',\omega}),
\end{equation}
which appear in the zero-temperature versions of
Eqs.~(\ref{2.22}) and (\ref{2.23}),
are convergent along any contour $\mathcal{C}$ running from
the origin (for metals the origin should be excluded, because
of the pole at that point) to infinity in a chosen direction
of the upper half-plane and yield always the
same value, provided that $\delta\varepsilon(\mathbf{r,\omega})$ 
approaches zero at least as $\mathcal{O}(1/|\omega|^{1+\delta})$ 
\mbox{($\delta$ $\!>$ $\!0$)} when $|\omega|$ goes to infinity,
which is the case.

Hence, the frequency integration
(for zero temperature) in Eqs.~(\ref{2.22}) and (\ref{2.23})
\mbox{($\tensor{G}$ $\!\mapsto$ $\!\tensor{G}^{\rm scat}$)}
can be performed along the positive imaginary axis (instead of
the positive real axis) and the imaginary part can be taken
after the integration has been performed. In this way,
Eq.~(\ref{2.35}) becomes equivalent to Eq.~(\ref{2.42}).   
To include the thermal weighting factor (which behaves
like $\omega^{-1}$ for \mbox{$\omega$ $\!\to$ $\!0$}), we first
note that integrands of the type which appear in in Eqs.~(\ref{2.22})
and (\ref{2.23})
\mbox{($\tensor{G}$ $\!\mapsto$ $\!\tensor{G}^{\rm scat}$)}
remain perfectly regular at \mbox{$\omega$ $\!=$ $\!0$}, thus
\begin{multline}
\label{C9}
\int_0^\infty d\omega \,
\coth\!\left(\frac{\hbar\omega}{2k_{\rm B} T}\right)
\Im\,[\ldots]
\\
= \lim_{\eta\to 0+} \Im \int_{\eta}^\infty d\omega \,
\coth\!\left(\frac{\hbar\omega}{2k_{\rm B} T}\right) \ldots .
\end{multline}
Since $\coth[\hbar\omega/(2k_{\rm B} T)]$ is holomorphic
and bounded for \mbox{$\Re \omega$ $\!\ge$ $\!\eta$},
we now may change the integration path in Eqs.~(\ref{2.22})
and (\ref{2.23}) according to
\begin{multline}
\label{C10}
\int_0^\infty d\omega \,
\coth\!\left(\frac{\hbar\omega}{2k_{\rm B} T}\right)
\Im\,[\ldots]
\\
=
\lim_{\eta\to 0+} \Im
\int_{\eta}^{\eta+i \infty} d\omega \,
\coth\!\left(\frac{\hbar\omega}{2k_{\rm B} T}\right)\cdots,
\end{multline}
which then leads to Eq.~(\ref{2.36}). 


\section{Derivation of Eq.~(\ref{2.46})}
\label{calculation1D}

The 1D electric field strength $\underline{\hat{\mathbf{E}}}
(\mathbf{r},\omega)$ as given in Eq.~(\ref{2.45}) implies
the equations
\begin{align}
\label{D1}
\underline{\hat{\mathbf{B}}}(\mathbf{r},\omega)
&=\frac{\mathbf{e}_y}{i\omega\sqrt{\mathcal{A}}}
\,\frac{\partial\underline{\hat{E}}(z,\omega)}{\partial z}\,, \\
\label{D2}
\underline{\hat{\mathbf{D}}}(\mathbf{r},\omega)
&=-\frac{\mathbf{e}_x}{\mu_0\omega^2\sqrt{\mathcal{A}}}
\,\frac{\partial^2 \underline{\hat{E}}(z,\omega)}{\partial z^2}
\end{align}
in place of Eqs.~(\ref{2.10}) and (\ref{2.11}),
so that Eq.~(\ref{2.17}) reduces to
\begin{equation}
\label{D3}
T_{zz}(\mathbf{r,r'},t)
=- {\textstyle\frac{1}{2}}
\bigl\langle\hat{\mathbf{D}}(\mathbf{r},t)
\hat{\mathbf{E}}(\mathbf{r'},t)
+\hat{\mathbf{B}}(\mathbf{r},t)
\hat{\mathbf{H}}(\mathbf{r'},t)\bigr\rangle.
\end{equation} 
We now proceed further as in the 3D case and
substitute the 1D version of Eq.~(\ref{2.12}) into
Eq.~(\ref{D3}), where the 1D Green function solves the
inhomogeneous wave equation
\begin{equation}
\label{D4}
\left[\frac{\partial^2}{\partial z^2}
+\frac{\omega^2}{c^2}\varepsilon(z,\omega)\right]G(z,z',\omega)
=-\delta(z-z').
\end{equation}
Restricting our attention to the zero-temperature limit,
we derive, on applying the 1D versions of Eqs.~(\ref{2.13b})
and (\ref{2.13c}),
\begin{eqnarray}
\label{D7}
\lefteqn{
\bigl\langle 0\bigr|\underline{\hat{\mathbf{D}}}(\mathbf{r},\omega)
\underline{\hat{\mathbf{E}}}{^\dagger}(\mathbf{r}',\omega')
\bigl| 0\bigr\rangle
}
&&\nonumber\\
&=&-\frac{\mu_0 \omega'^2}{\mathcal{A}}
\frac{\partial^2}{\partial z^2}
\bigl\langle 0\bigr|(G \star W\hat{f})(z,\omega)
(W\hat{f}^\dagger \star G^\ast)(z',\omega')\bigl|0\bigr\rangle
\nonumber\\
&=&-\frac{\hbar}{\pi\mathcal{A}}
\,\delta(\omega-\omega')\,\frac{\partial^2}{\partial z^2}
\frac{\omega^2}{c^2}(G\star \varepsilon''G^\ast)(z,z',\omega)\nonumber\\
&=&-\frac{\hbar}{\pi\mathcal{A}}\delta(\omega-\omega')
\frac{\partial^2}{\partial z^2}\,\Im\left[ G(z,z',\omega)\right]
\nonumber\\
&=&\frac{\hbar}{\pi\mathcal{A}}
\,\delta(\omega-\omega')\,
\frac{\omega^2}{c^2}\,\Im\left[\varepsilon(z,\omega)G(z,z',\omega)\right]
\end{eqnarray}
and
\begin{eqnarray}
\label{D8}
\lefteqn{
\bigl\langle 0\bigr|\underline{\hat{\mathbf{B}}}(\mathbf{r},\omega)
\underline{\hat{\mathbf{H}}}{^\dagger}(\mathbf{r}',\omega')
\bigl|0\bigr\rangle
}&&\nonumber\\
&=&\frac{\mu_0 \omega\omega'}{\mathcal{A}}\!
\frac{\partial^2}{\partial z \partial z'}
\bigl\langle 0\bigr|(G \star W\hat{f})(z,\omega) (W\hat{f}^\dagger \star
G^\ast)(z',\omega')\bigl|0\bigr\rangle
\nonumber\\
&=&\frac{\hbar}{\pi\mathcal{A}}
\,\delta(\omega-\omega')\,
\frac{\partial^2}{\partial z \partial z'}
\frac{\omega^2}{c^2}(G\star \varepsilon''G^\ast)(z,z',\omega)
\nonumber\\
&=&\frac{\hbar}{\pi\mathcal{A}}
\,\delta(\omega-\omega')\,
\frac{\partial^2}{\partial z \partial z'}
\,\Im\left[ G(z,z',\omega)\right]
\end{eqnarray}
[for $W$ and the $\star$-notation, see Eqs.~(\ref{A10})
and (\ref{A11})]. By Fourier transforming the fields back
into the time domain and combining with Eq.~(\ref{D3}), 
we arrive at
\begin{multline}
\label{D9}
T_{zz}(\mathbf{r,r'})
=-\frac{\hbar}{2\pi\mathcal{A}}\int_0^\infty d\omega
\ \times \\ \ \times
\Im\left[\frac{\omega^2}{c^2}\varepsilon(z,\omega)G(z,z',\omega)
+\frac{\partial^2}{\partial z \partial z'}G(z,z',\omega)\right].
\end{multline}
Note that Eq.~(\ref{D9}) is valid for an arbitrary
$\varepsilon(z,\omega)$. 

Let us now specify $\varepsilon(z,\omega)$ according
to Eq.~(\ref{2.15}). The bulk Green function in the $j$th layer
reads
\begin{eqnarray}
\label{D5}
&&
G_j^{\rm bulk}(z,z',\omega)=\frac{i}{2\beta_j}
\,e^{i\beta_j|z-z'|}
\nonumber\\&&\hspace{1ex}
= g_j^{(0)+} e^{i\beta_j z} + g_j^{(0)-} e^{-i\beta_j z}
\nonumber\\&&\hspace{1ex}
=\frac{i}{2\beta_j}[e^{i\beta_j(z-z')}\theta(z\!-\!z')
+e^{-i\beta_j(z-z')}\theta(z'\!-\!z)]
\qquad
\end{eqnarray}
[$\theta(z)$, Heaviside step function], where the propagation constant
\begin{equation}
\label{D6}
\beta_j = \frac{\omega}{c}\sqrt{\varepsilon_j(\omega)}
\end{equation}
coincides with the full wave number. The scattering part of the
Green function in the $j$th layer
\begin{equation}
\label{D10}
G^{\rm scat}_j(z,z',\omega)
= g_j^+ e^{i\beta_j z} + g_j^- e^{-i\beta_j z}
\end{equation} 
is the solution of the homogeneous wave equation,
where the matching conditions
\begin{gather}
\label{D11}
g_{j}^+
=r_{j-} \Bigl(g_{j}^{(0)-}+g_{j}^-\Bigr)
\quad (z = 0 < z'),
\\
\label{D12}
g_{j}^- e^{-i\beta_j d_{j}}
=r_{j+}e^{i\beta_j d_{j}}
\Bigl(g_{j}^{(0)+}+g_{j}^+\Bigr)
\quad (z = d_j > z')
\end{gather}     
must be satisfied. Thus
\begin{gather}
\label{D13}
g_j^\pm=\frac{r_{j\mp}}{D_{j}}\,e^{i\beta_j d_{j}}
\Bigl(
e^{\mp i\beta_j d_{j}}g_{j}^{(0)\mp}+e^{i\beta_j d_{j}}
r_{j\pm} g_{j}^{(0)\pm} \Bigr),
\\
\label{D14}
D_{j}\equiv 1-r_{j+} r_{j-}e^{2i\beta_j d_{j}}.
\end{gather}
The 1D reflection coefficients $r_{j\pm}$ can be constructed
as outlined in Appendix \ref{Fresnel}. Obviously, they are
identical to the 3D ones (for $s$-polarization)
taken at \mbox{$\mathbf{q}$ $\!=$ $\!0$} \footnote{Although
   distinguishing the two polarizations is meaningless at
   \mbox{$\mathbf{q}$ $\!=$ $\!0$}, one should use the
   $s$-coefficients. The $p$-coefficients differ in sign.}.
   Combining Eqs.~(\ref{D10}) and (\ref{D13}), we find that,
on taking $g_{j}^{(0)\pm}$ from Eq.~(\ref{D5}),
\begin{eqnarray}
\label{D15}
&&
G^{\rm scat}_j(z,z',\omega)
=\frac{ie^{i\beta_jd_{j}}}{2\beta_jD_j}
\Bigl[r_{j-}e^{-i\beta_jd_{j}}e^{i\beta_j(z+z')}
\nonumber\\&&\hspace{2ex}
+\,r_{j-}r_{j+}e^{i\beta_jd_{j}}e^{i\beta_j(z-z')}
+r_{j+}e^{i\beta_jd_{j}}e^{-i\beta_j(z+z')}
\nonumber\\&&\hspace{2ex}
+\,r_{j+}r_{j-}e^{i\beta_jd_{j}}e^{-i\beta_j(z-z')}\Bigr].
\end{eqnarray}
From Eq.~(\ref{D15}) it then follows that
\begin{multline}
\label{D16}
\frac{\omega^2}{c^2}\varepsilon(z,\omega)
G^{\rm scat}_j(z,z',\omega)
+\frac{\partial^2G^{\rm scat}_j(z,z',\omega)}{\partial z\partial z'}
\\
=\frac{2i\beta_jr_{j+}r_{j-}e^{2i\beta_jd_{j}}}{D_j}
\cos[\beta_j(z-z')].
\end{multline}
Substituting this expression into Eq.~(\ref{D9}) [$G(z,z',\omega)$
$\!\mapsto$ $\!G^{\rm scat}(z,z',\omega)$] and setting \mbox{$z'$
$\!=$ $\!z$},
we eventually arrive at
Eq.~(\ref{2.46}).


\section{Derivation of Eq.~(\ref{2.80})}
\label{Evaluation}

We first note that Eq.~(\ref{2.78}) can be rewritten as
\begin{equation}
\label{E1}
F
\approx \frac{\hbar}{8 \pi^2 d_{j}^3}\int_0^\infty d\xi\, 
{\rm Li}_3\!
\left[\frac{(\varepsilon_{j+1} -1)(\varepsilon_{j-1} -1)}
{(\varepsilon_{j+1}+1)(\varepsilon_{j-1} +1)}\right].
\end{equation} 
From Eq.~(\ref{2.79}) it follows that
\begin{equation}
\label{E2}
\frac{\varepsilon_{j\pm 1}(i\xi)-1}{\varepsilon_{j\pm 1}(i\xi)+1}
=\frac{1}{1+2[\xi(\xi+\gamma_0)+\omega_0^2]/\Omega^2}\,.
\end{equation}
Substituting this expression into Eq.~(\ref{E1}), we obtain
\mbox{[$y$ $\!=$ $\!(\xi$ $\!+$ $\!\gamma_0/2)/\Omega$]}
\begin{equation}
\label{E3}
F
\approx \frac{\hbar\Omega}{8\pi^2 d_{j}^3}
\sum_{m=1}^\infty \frac{I_m}{m^3}\,,
\end{equation}
where
\begin{equation}
\label{E3a}
I_m = \int _{\frac{\gamma_0}{2\Omega}}^\infty dy \,
\left[
1+2(y^2+\alpha^2)
\right]^{-2m}
\end{equation}
and
\begin{equation}
\label{E4}
\alpha^2\equiv \frac{\omega_0^2-\gamma_0^2/4}{\Omega^2}.
\end{equation}
Under the condition that $\gamma_0$ is
sufficiently small, so that
\mbox{$\gamma_0$ $\!\ll$ $\!2\Omega$} and \mbox{$\gamma_0^2/4$ $\!<$ $\!\omega_0^2$},
the integral (\ref{E3a}) is approximately given by
\begin{equation}
\label{E5}
I_m = \frac{1}{2^{2m+1}}\int _{-\infty}^\infty 
\frac{dy}{\left({y^2+\alpha^2+1/2}\right)^{2m}}\,,
\end{equation}
which can be easily evaluated by employing the
residue theorem. For this purpose we write

\begin{eqnarray}
\label{E7}
\lefteqn{
\frac{1}{\left(y^2+\alpha^2+1/2\right)^{2m}} \nonumber
}
\nonumber\\&&
= - \frac{2m}{(2y_{+})^{2m}}
\left[
\frac{1}{(y-y_{+})(y-y_{-})^{2m-1}} \right.
\nonumber\\&&\hspace{3ex}
\left.
+\,\frac{1}{(y-y_{-})(y-y_{+})^{2m-1}}
\right]+\cdots,
\end{eqnarray}
where
\begin{equation}
\label{E6}
y_{\pm}=\pm i \sqrt{\alpha^2+1/2}\,.
\end{equation}
We thus derive
%
\begin{equation}
\label{E8}
I_{m}=\frac{\pi}{2^{6m-1}}\frac{\Gamma(4m-1)}
{[\Gamma(2m)]^2}\,{(\alpha^2+1/2)^{1/2-2m}}\,.
\end{equation}
%
%
Combining Eqs.~(\ref{E3}) and (\ref{E8}) eventually yields
Eq.~(\ref{2.80}). Because of
\begin{multline}
\label{E9}
\sum_1^\infty\frac{1}{m^3}\int_0^{\gamma_0/2\Omega} dy\,
\left[1+2(y^2+\alpha^2) \right]^{-2m} \\
\le
\frac{\gamma_0}{2\Omega}{\rm Li}_3\!
\left[(\alpha^2+1/2)^{-2}/4 \right],
\end{multline}
the small relative error that results from replacing the lower limit of integration
in Eq.~(\ref{E3a}) by zero can be estimated according to
\begin{equation}
\label{E10}
\frac{\delta F}{F} \le \frac{\gamma_0}{\Omega}\, f(\alpha^2+1/2),
\end{equation}
where
[see Eq.~(\ref{tildedilog})]
%
\begin{equation}
\label{E11}
f(x)=\frac{1}{8\pi\sqrt{x}}\, \frac{{\rm Li}_3(x^{-2}/4)}
{\widetilde{\rm Li}_2(x^{-2}/64)}\,.
\end{equation}
Note that the factor \mbox{$f(\alpha^2$ $\!+$ $\!1/2)$} is at most
of order of unity.
\end{appendix}


\end{document}